\documentclass[journal=jacsat,manuscript=article]{achemso}

\usepackage[version=3]{mhchem} 
\usepackage{booktabs}
\usepackage{threeparttable}
\usepackage{siunitx}
\usepackage{bm}

\author{Alex Glover}
\author{Maria M. Papathanasiou}
\email{maria.papathanasiou11@imperial.ac.uk}
\author{Ronny Pini}
\email{r.pini@imperial.ac.uk}
\affiliation[Imperial]
{Department of Chemical Engineering, Imperial College London \\
The Sargent Centre for Process Systems Engineering, Imperial College London}

\title[An \textsf{achemso} demo]
  {Accelerating Simulation and Optimisation of Cyclic Adsorption Processes with Differentiable Programming}

\abbreviations{IR,NMR,UV}

\begin{document}







\begin{abstract}

The design of cyclic adsorption processes is computationally demanding, requiring repeated convergence to cyclic steady state within an iterative optimisation loop. Conventional workflows treat the process simulator as a black box and rely on derivative-free optimisation, resulting in design campaigns that can require hundreds to thousands of CPU hours. This work presents an end-to-end differentiable model of a pressure vacuum swing adsorption process, developed using the JAX differentiable programming framework and applied here to a benchmark post-combustion carbon capture problem. Automatic differentiation provides exact gradients throughout the entire computational workflow. The differentiation of a single process cycle provides the Jacobian for a Newton iteration to decrease both the number of iterations and the simulation time required to reach cyclic steady state by a factor of 20 relative to a representative MATLAB implementation. Exact gradients of the performance metrics with respect to the design variables further enable gradient-based multi-objective optimisation using the IPOPT algorithm. Applied to a six-variable design problem, the latter produces a superior Pareto front with improved coverage of the trade-off space and closer convergence to the optimal front than the genetic algorithm NSGA-II. Notably, the full front is obtained over two orders of magnitude faster than the conventional approach. By retaining the full mechanistic model while making it differentiable, this framework transforms cyclic adsorption process design from slow black-box simulation with derivative-free optimisation to efficient gradient-enhanced modelling and optimisation, enabling rapid and systematic exploration of complex design spaces.

\end{abstract}

\section{Introduction}\label{Introduction}

Adsorption-based processes are widely employed in diverse gas separation applications and have received increasing interest for post-combustion and other carbon dioxide capture technologies.\cite{raganati_adsorption_2021, pini_co2_2026} The process itself is operated through a sequence of steps, in which either the pressure or the temperature, or both, are varied to selectively adsorb and desorb the components in the gas mixture. Mechanistic adsorption models, such as the one-dimensional axially dispersed plug flow model, have been used extensively throughout the literature for experiment design and interpretation, and to describe the behaviour observed in cyclic processes.\cite{ward_integrated_2022, krishnamurthy_co2_2014, todd_fast_2001} These models have been used for open-loop predictions to screen adsorbents \cite{rajagopalan_adsorbent_2016} and for integrated adsorbent-process optimization.\cite{khurana_integrated_2017} The design exercise is not trivial and requires finding the set of operating parameters (e.g., the duration and operating pressure of each step in the cycle) that balance conflicting objectives (or key performance indicators, KPIs) such as the purity and recovery of the product as well as the productivity and energy consumption of the process. This design workflow is typically formulated as a multi-objective optimisation problem that incorporates a combination of these four KPIs, or as a single-objective economic optimisation.\cite{ward_efficient_2022, haghpanah_multiobjective_2013} Recent years have seen developments that greatly expand the design space which needs to be explored, including novel adsorbent materials (e.g., metal organic frameworks \cite{mahajan_recent_2022}), multi-sorbent beds \cite{ward_design_2024}, structured contactors \cite{antunes_morgado_3d_2026} as well as process configurations featuring additional steps beyond the foundational 2-bed, 4-step Skarstrom cycle \cite{pai_superstructure_2025}. The optimisation of these more advanced designs increases the computational demand, with individual problems often requiring hundreds to thousands of CPU hours.\cite{ward_efficient_2022, peh_metal-organic_2022, wilkins_pilot-scale_2025, subraveti_machine_2019, pai_experimentally_2020} 

The ability to accurately determine model gradients underpins several applications of the mechanistic adsorption model just described – from sensitivity analysis to optimisation. This need becomes critical as the total number of parameters increases and when dealing with highly non-linear processes, as it is the case for the design of adsorption-based separations. Differentiable programming, a framework in which the numerical programme is constructed such that it can be \emph{automatically} differentiated throughout (i.e. without manual derivation), offers a route to drastically increase the computational efficiency of these tasks. The benefits of differential programming have been demonstrated in several application areas, including crystallisation,\cite{alsubeihi_modern_2025} computational fluid dynamics,\cite{bezgin_jax-fluids_2023} chemical kinetics,\cite{lee_differentiable_2025} and molecular dynamics.\cite{schoenholz_jax_2021} In this work, we develop a differentiable framework for the simulation and optimisation of cyclic adsorption processes. A one-dimensional axially dispersed plug flow model is implemented in JAX, leveraging just-in-time compilation, automatic differentiation, and accelerated linear algebra throughout the process model. These capabilities are used to accelerate the numerical solution, the convergence to cyclic steady state, and the multi-objective optimisation. The following subsections review the modelling and optimisation approaches used within the adsorption community and contrast these with methods from the process systems engineering literature, highlighting the gap that differentiable programming can address.

\subsection{Adsorption Process Design}

The physical phenomena in the fixed-bed adsorption column are described by mass and energy balances alongside equations characterising the local adsorption behaviour, yielding a system of partial differential equations (PDEs). A method of lines approach, typically using a finite volume scheme, is often used to discretise the system of PDEs into a set of ordinary differential equations (ODEs).\cite{haghpanah_multiobjective_2013} Upon application of the relevant boundary conditions, these equations are solved for each step of the adsorption process using an implicit solver due to the stiffness of the system.\cite{leveque_finite_2002} Owing to the repeated sequence of steps imposed on the adsorption column, the process is designed to operate at cyclic steady state (CSS) -- an operating point for which the state of the column at any given time matches its state at the exact same time in the previous cycle. Reaching CSS typically involves simulation from a startup condition (a clean or saturated bed) through repeated cycles until periodic stability is achieved. Adsorption processes with short cycle times and pressure rather than temperature swings may require hundreds of cycles to reach CSS.\cite{croft_periodic_1994}
 
The optimisation of adsorption processes has predominantly been carried out using derivative-free or black box methods due to their simplicity and their ability to deal with non-convex optimisation problems. Here, the cyclic process model is iterated until CSS is reached; the objective function and performance constraints (typically the KPIs) are then evaluated, and the optimisation algorithm computes the next set of parameters for evaluation. This structure is displayed in Figure \ref{optimisation structures}a. A genetic algorithm, such as the non-dominated sorting genetic algorithm II (NSGA-II),\cite{deb_fast_2000} is by far the most widely used derivative-free approach in the adsorption literature\cite{haghpanah_multiobjective_2013, wilkins_pilot-scale_2025, subraveti_machine_2019, ward_design_2024, pai_generalized_2020, beck_efcient_2014, liao_simulation_2025}. In practice, this approach requires simulating thousands of candidate designs, each of which must be iterated to CSS, resulting in the large computational times reported in the literature.\cite{ward_efficient_2022, peh_metal-organic_2022, wilkins_pilot-scale_2025, subraveti_machine_2019, pai_experimentally_2020} Other derivative-free approaches, such as Bayesian optimisation, have been also applied, demonstrating reduced computational times relative to genetic algorithms.\cite{hao_efficient_2021, ward_efficient_2022} Alternatively, many studies use simplified models\cite{khurana_adsorbent_2016, subramanian_balashankar_analysis_2019} or machine learning surrogates trained from a set of high-fidelity simulations\cite{subraveti_machine_2019, chatziasteriou_surrogate_2025} to reduce the computational cost which are often formulated in a way that alleviates the requirement of iterating to CSS.

\subsubsection{Gradient-based methods for cyclic process optimisation}

The process systems engineering literature reports on a range of gradient-based methods for the optimisation of periodic processes that address many of the computational limitations described above, though these have seen limited adoption in adsorption applications. Early approaches accelerated convergence to CSS using Newton or quasi-Newton iterations applied to the direct determination of the periodic state.\cite{croft_periodic_1994, ding_periodic_2001} These methods achieve rapid (quadratic) convergence, but require evaluation of the Jacobian matrix to describe the sensitivities of the final state of the cycle with respect to the initial state. While convergence to CSS is achieved in much fewer iterations than successive substitution, the additional computational burden of the Jacobian evaluation has limited their adoption.  \cite{ko_optimization_2003, ding_periodic_2001, jiang_simulation_2003} Jacobian-free vector-based extrapolation methods have been proposed\cite{subraveti_study_2024, subraveti_vector_2026}, achieving up to fourfold speed-ups while remaining simple to implement. However, their convergence rate remains inferior to exact Newton iteration methods. Machine learning models have also been used to predict approximate CSS profiles, providing improved initial guesses that accelerate convergence in subsequent simulations \cite{pai_experimentally_2020}.
 
\begin{figure}
    \centering
    \includegraphics[width=1\linewidth]{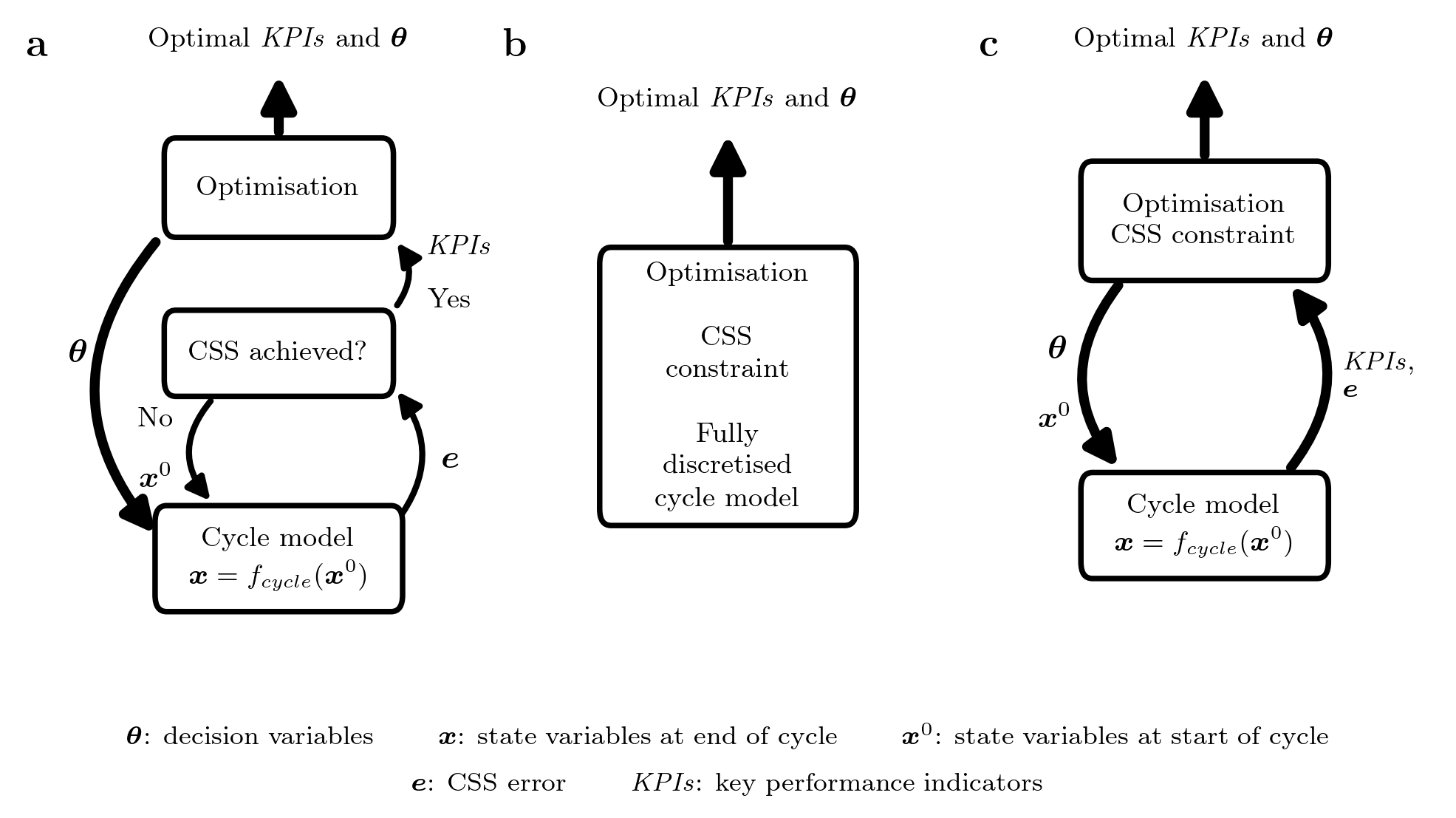}
    \caption{Optimisation structures used within the literature: derivative-free optimisation (a), complete discretisation (b), and simultaneous tailored approach (c). Adapted from \citeauthor{biegler_recent_2005}\cite{biegler_recent_2005}}
    \label{optimisation structures}
\end{figure}

An alternative strategy for the optimisation of periodic processes is full discretisation of both space and time \cite{nilchan_optimisation_1998}, transforming the governing PDEs into a nonlinear algebraic system of equations solved via nonlinear programming. While it enables direct inclusion of the CSS criterion and the optimisation objectives (Figure \ref{optimisation structures}b), this approach is highly sensitive to the initial guess and can suffer from convergence issues for complex process models.\cite{ko_optimization_2003, jiang_simulation_2003} 
A pseudo-transient formulation has been proposed to mitigate this by gradually introducing system dynamics, improving robustness\cite{tsay_pseudo-transient_2018}. 
 
In practice, the single discretisation approach -- where PDEs are discretised only in space and the resulting ODEs are integrated in time -- is more commonly used. Within this framework, simultaneous optimisation formulations incorporate the CSS convergence criterion as an inequality constraint (Figure \ref{optimisation structures}c).\cite{jiang_simulation_2003} Although it avoids explicit iteration to CSS, this approach increases problem size and complexity, as both the state variables at every discretised point and the design variables become decision variables. These formulations rely on gradient-based algorithms, requiring sensitivity information that can be obtained via finite differences,\cite{cruz_optimization_2005} adjoint ODEs,\cite{latifi_optimisation-based_2008} or through discrete methods.\cite{jiang_simulation_2003} Many of these optimisation problems have been solved using the reduced Hessian Sequential Quadratic Programming (SQP) algorithm.\cite{ternet_recent_1998, biegler_recent_2005} 
Other works have formulated the design problem as an optimal control problem using dynamic optimisation approaches.\cite{agarwal_superstructure-based_2010, dowling_large-scale_2012, khajuria_optimization_2013} 

The computation of sensitivities is typically the dominant computational cost in the gradient-based approaches described above, often exceeding 90\% of the CPU time,\cite{jiang_simulation_2003, kawajiri_optimization_2006} and motivating the development of optimisation algorithms that can accommodate inexact Jacobians.\cite{vetukuri_inexact_2010} Despite these advances, the need for accurate sensitivities -- coupled with approximation errors, stability concerns, or implementation effort -- has hindered broader adoption. Automatic differentiation within a differentiable programming framework offers exact, stable sensitivities without manual derivation. While differentiable programming has been successfully applied in adjacent process engineering domains,\cite{alsubeihi_modern_2025, bezgin_jax-fluids_2023, lee_differentiable_2025, schoenholz_jax_2021} its application to the rigorous simulation and optimisation of cyclic adsorption processes has not been demonstrated.

\section{A differentiable framework for adsorption processes}

In this section, we introduce the differentiable framework for the simulation and optimisation of cyclic adsorption processes. The framework follows the same modelling approach used throughout the adsorption literature: the governing PDEs are discretised with a finite volume scheme, the resulting ODEs are integrated with an implicit solver for each step of the cycle, and the process is iterated to CSS from which the KPIs are computed for use within a multi-objective optimisation problem. The mechanistic model is therefore unchanged. The new development is in making the entire computational workflow differentiable, enabling the use of gradient information for the accelerated simulation and optimisation of the adsorption process.

\begin{figure}[h!]
    \centering
    \includegraphics[width=0.8\linewidth]{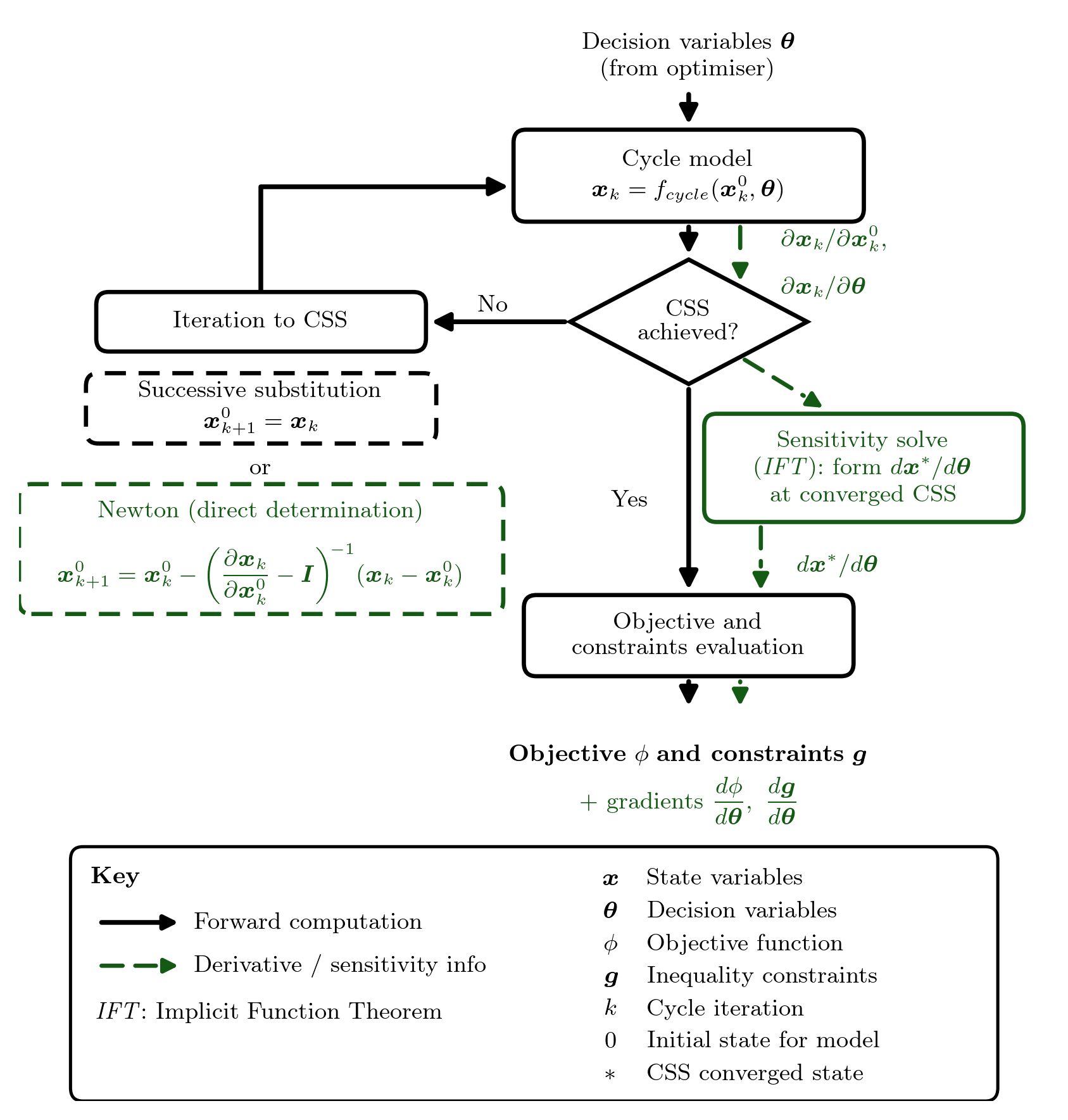}
    \caption{Differentiable adsorption model framework. The sensitivity propagation for the computation of the gradients of the objective functions (e.g., KPIs) and the constraints with respect to the decision variables are shown with the green arrows. The two options for the iteration to CSS are illustrated, namely successive substitution (dashed black box) and direct determination (dashed green box).}
    \label{differentiable model framework}
\end{figure}
 
As identified in the preceding section, the conventional model formulation in the adsorption literature does not exploit gradient information at any stage of the design process. The differentiable framework addresses this at three levels, as illustrated in Figure \ref{differentiable model framework}. First, at the level of the ODE solution, just-in-time (JIT) compilation and automatic differentiation (AD) within the implicit solver are used to accelerate the numerical integration and provide sensitivities of the solution with respect to the initial state and model parameters (green dashed arrows in the figure). Second, these sensitivities enable a Newton-based direct determination method for convergence to CSS, replacing the successive substitution approach (dashed green box in the figure). Third, the CSS condition is treated as an implicit function, such that the model outputs both objective functions (e.g., KPIs) and their gradients with respect to the decision variables for use within gradient-based multi-objective optimisation. These three components are described in the following subsections.

\begin{figure}[h!]
    \centering
    \includegraphics[width=0.8\linewidth]{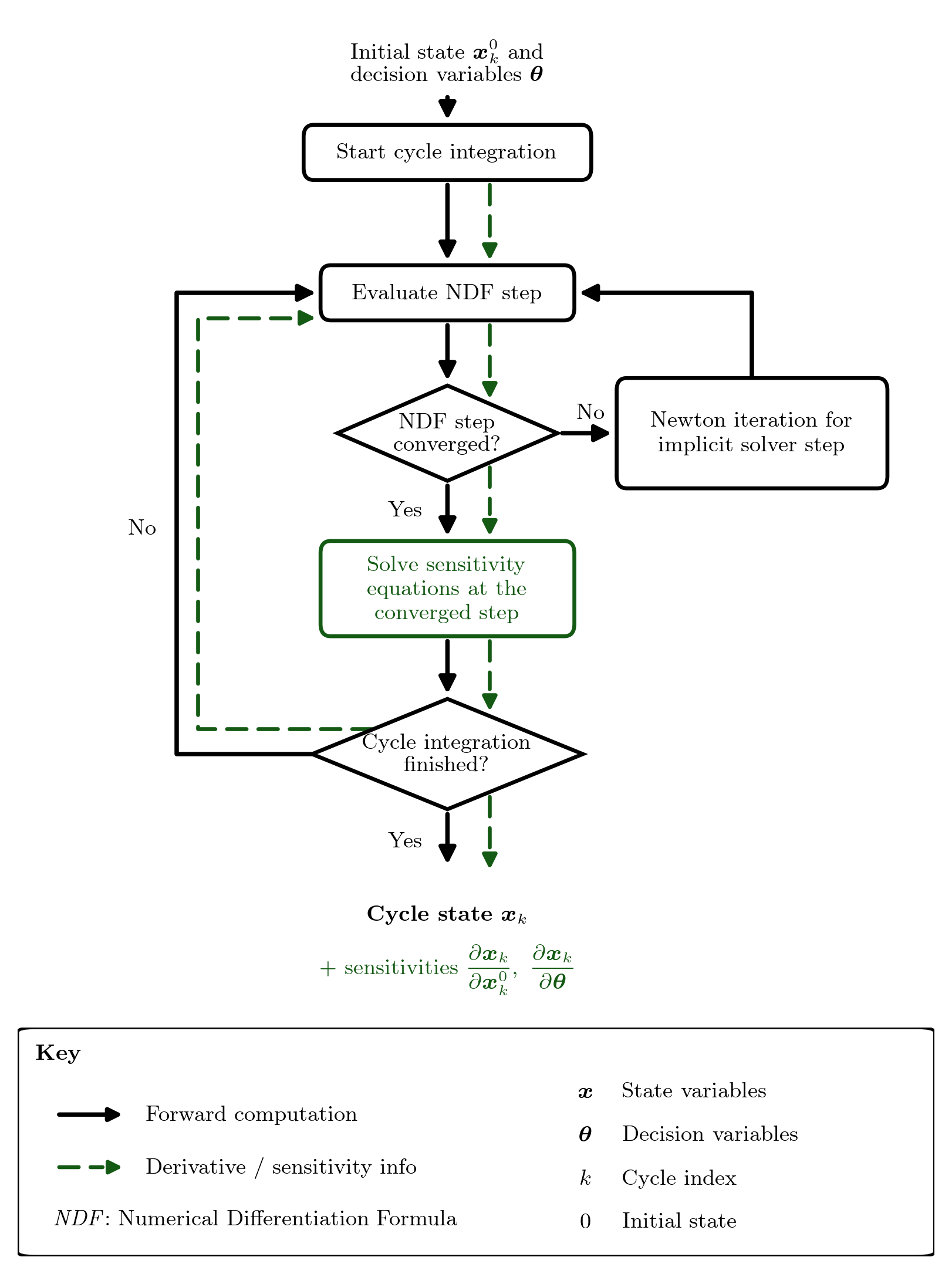}
    \caption{Flowchart for the state and the sensitivities computation during the solution of one adsorption cycle. }
    \label{fig: cycle solution and sensitivities}
\end{figure}

\subsection{Solution and Sensitivities}\label{Differentiable Programming}

The system of PDEs that describes the adsorption process (Table \ref{pde equations}) is discretised via a finite volume scheme where the fluxes at the walls are evaluated using a third-order weighted essentially non-oscillatory method.\cite{haghpanah_multiobjective_2013} The resulting dimensionless discretised equations are given in the Section S2 of the Supporting Information. Equation \ref{general ode} describes their general form, where $\textbf{x}$ denotes the vector of dimensionless state variables (pressure, temperature, mole fractions, column wall temperature and adsorbed quantities) at each volume node along the column, $\boldsymbol{\theta}$ denotes the model parameters, and $\textbf{x}_{0}$ is the state of the column at the start of the step. 

\begin{equation}\label{general ode}
    \frac{d\textbf{x}}{dt}= f(t,\textbf{x}(t),\boldsymbol{\theta}),
    \qquad
    \textbf{x}(0) = \textbf{x}_{0}
\end{equation}

Figure \ref{fig: cycle solution and sensitivities} illustrates the flowchart for the solution of one process cycle as well as the sensitivities computation and propagation through the model solution. An implicit ODE solver is chosen for the integration of the states due to the stiffness of the equations resulting from the discretised second order spatial derivatives (diffusion and conduction) and the adsorption and heat transfer source/sink terms.\cite{leveque_finite_2002} At each time step of the implicit solver, a system of non-linear equations must be solved, typically via Newton iterations, which in turn requires the evaluation of the Jacobian of the right-hand side function $f(t,\textbf{x}(t),\boldsymbol{\theta})$ with respect to the state vector $\textbf{x}$. With the differentiable framework adopted here, this Jacobian is computed using AD rather than being approximated with finite differences or derived manually. 

A key requirement for the gradient-based methods used in this work is the ability to obtain sensitivities of the solution with respect to the initial state ($\partial \textbf{x}_k/\partial \textbf{x}^0_k$) and model parameters ($\partial \textbf{x}_k/\partial \bm{\theta}$) -- see Figure~\ref{fig: cycle solution and sensitivities}. For implicit solvers, such as the one used in this work, a naive use of AD would differentiate through all Newton iterations at each time step, leading to high computational cost and numerical sensitivity to the solver convergence tolerance. Instead, the sensitivities are computed \emph{at convergence} of the non-linear system as illustrated in Figure \ref{fig: cycle solution and sensitivities}. This task is achieved by directly differentiating the numerical time-stepping scheme using the discrete sensitivity method, implemented in the forward direction.\cite{sapienza_differentiable_2024} The resulting tangent gradients are then ``manually" propagated alongside the state variables throughout the simulation. To enable comparison with conventional approaches used in adsorption modelling, we have implemented in JAX a version of the numerical differentiation formula (NDF) method used in MATLAB's \texttt{ode15s} solver\cite{shampine_matlab_1997}. The corresponding state and sensitivity equations are given in Section S1 of the Supporting Information. Notably, in contrast to \texttt{ode15s}, the Jacobian is re-evaluated at every \emph{accepted} step to ensure accurate sensitivity calculations, rather than being reused at many steps when sensitivities are not required.

The Jacobian matrices required in both the state and sensitivity equations exhibit a known sparsity pattern arising from the spatial structure of the discretised transport equations for the adsorption column. The sparsity is exploited by partitioning the right-hand side function $f(t,\textbf{x}(t),\boldsymbol{\theta})$ so that each evaluation operates only on the subset of volumes that directly influence a given spatial location. AD is then applied locally to these reduced functions, computing only the non-zero derivative entries. These operations are vectorised across all control volumes using JAX's \texttt{vmap} function. The linear systems arising in both the Newton iterations and the sensitivity equations share the same sparsity pattern and are solved using a sparse linear solver (KLU algorithm \cite{noauthor_gdsfactoryklujax_2026}). This exploitation of sparsity ensures that the approach scales to finer spatial discretisation.

\subsection{Direct Determination of Cyclic Steady State}\label{direct determination}

Adsorption processes are typically cyclic in nature where each cycle is composed of a sequence of steps for which the end state of a cycle becomes the initial state of the following cycle. In modelling studies, a mathematical criterion is used to define whether a process is operating at CSS. Throughout the literature, authors have enforced this criterion based upon change in the values of the state variables\cite{nilchan_optimisation_1998}, KPIs\cite{marcinek_performance_2020} and the overall mass balance for a cycle.\cite{krishnamurthy_co2_2014} In this work we define the cycle error as the difference between the vector of state variables at the start and end of a cycle:

\begin{equation}\label{error}
    \textbf{e}_{k} = \textbf{x}_{k} - \textbf{x}_{k}^{0}
\end{equation}

The CSS criterion is satisfied when all elements of this vector, normalised to the initial state $\textbf{x}_{k}^{0}$, are below a predefined tolerance (for example, $10^{-5}$). Typically, the so-called successive substitution method is used to reach the CSS. This method involves simulating the process from an initial condition (a bed saturated with the light or heavy component) to CSS. Mathematically, this is known as a Picard iteration where the output of the cycle function is used as the input to the next cycle. This method has the benefit of following the dynamics of the process from startup to CSS – information not necessarily required for optimisation studies where the target is a design at CSS. This method can be slow, especially for systems where the cycle time is short, which is common in pressure swing processes. The slow convergence is described by the CSS having a spectral radius, the largest eigenvalue of the cycle Jacobian matrix $\partial\textbf{x}_{k}/{\partial\textbf{x}_{k}^{0}}$ and also known as the dominant Floquet multiplier, close to, but less than 1 \cite{douglas_levan_determination_1995}. In contrast, the direct determination method does not follow the process dynamics; rather it predicts the input to the next cycle from a Newton iteration step\cite{croft_periodic_1994}:

\begin{equation}\label{DD Newton}
    \textbf{x}_{k+1}^{0} = \textbf{x}_{k}^{0} - \alpha\frac{\partial\textbf{e}_{k}}{\partial\textbf{x}_{k}^{0}}^{-1}\textbf{e}_{k}
\end{equation}

where the error, $\textbf{e}_{k}$, and the error gradient with respect to the initial state,  $\partial\textbf{e}_{k}/\partial\textbf{x}_{k}^{0}$, are computed at each iteration $k$. Figure \ref{differentiable model framework} illustrates the flowchart for the differentiable model, where the iteration to CSS is shown with the loop. The traditional successive substitution and the direct determination methods are shown. The error gradient is computed as follows: 

\begin{equation}\label{error gradient}
    \frac{\partial\textbf{e}_{k}}{\partial\textbf{x}_{k}^{0}} = \frac{\partial\textbf{x}_{k}}{\partial\textbf{x}_{k}^{0}} - I
\end{equation}

Again, the differentiable framework described above allows for the efficient evaluation of the full Jacobian matrix for the
sensitivities of the states at the end of the cycle with respect to the initial states at each iteration. As the Newton iteration does not follow the process dynamics, it can generate initial conditions for the model that include non-physical values,\cite{croft_periodic_1994} such as negative mole fractions or adsorbed quantities. To prevent failure of the ODE solver, a damped Newton iteration is implemented in Equation~\ref{DD Newton}, which uses a damping factor $\alpha \in [0,1]$ to scale the Newton step:

\begin{equation}\label{damping}
\alpha = \frac{1}{2} min( \frac{ x^{0}_{k,i} }{ x^{0}_{k,i} - x^{0,u}_{k+1,i}} )
\end{equation}

where $x^{0,\mathrm{u}}_{k+1,i}$ is the state variable for the next iteration from the undamped Newton iteration ($\alpha=1$), $i$ corresponds to the discretised state variable where negative values are encountered, and $x^{0}_{k,i}$ is the corresponding discretised state variable from the previous iteration. The minimum is hence only taken over state variables where negative values are encountered and the factor of $\frac{1}{2}$ provides a safety margin to ensure the next iterate is not close to negative values.

\subsection{Gradient-Based Optimisation}\label{gradient-based optimisation}

The gradient-based optimisation methodology uses the same structure as the derivative-free optimisation studies illustrated in Figure \ref{optimisation structures}a, where the iteration to CSS is the inner loop within each evaluation. Without loosing generality and with reference to Figure~\ref{differentiable model framework}, Equation \ref{general optimisation problem} describes an optimisation problem involving a single objective function $\phi$, alongside a set of inequality constraints $\textbf{g}$, which are both typically described through the evaluation of the KPIs of the process. The decision variables for the optimisation $\boldsymbol{\theta}$ are bounded (the subscript $L$ and $U$ refer to the lower and upper bound, respectively), and $\textbf{x}^{*}$ is the vector of the state variables at the start of the cycle where the CSS criterion is satisfied. 

\begin{equation}\label{general optimisation problem}
    \begin{aligned}
    \min_{\boldsymbol{\theta} } \quad & \phi(\textbf{x}^{*}, \boldsymbol{\theta})\\
    \textrm{s.t.} \quad & \textbf{g}(\textbf{x}^{*}, \boldsymbol{\theta}) \le 0\\
      & \boldsymbol{\theta}\in[\boldsymbol{\theta}^{L},\boldsymbol{\theta}^{U}]    \\
    \end{aligned}
\end{equation}

The CSS condition is expressed as an implicit function: 

\begin{equation}\label{CSS satisfied}
    \textbf{x}^{*} - \textbf{f}_{cycle}(\textbf{x}^{*}, \boldsymbol{\theta}) = 0
\end{equation}

where the function $\textbf{f}_{cycle}$ describes the whole process cycle. At each iteration, the gradients of the objective function and the constraints with respect to the design parameters are computed as follows:

\begin{equation}\label{objective function gradients}
    \frac{d\phi}{d \boldsymbol{\theta} } = \frac{\partial \phi}{\partial \boldsymbol{\theta} } + \frac{\partial \phi}{\partial \textbf{x}^{*} } \frac{d\textbf{x}^{*}}{d \boldsymbol{\theta} }
\end{equation}

\begin{equation}\label{constraint gradients}
    \frac{d\textbf{g}}{d \boldsymbol{\theta} } = \frac{\partial \textbf{g}}{\partial \boldsymbol{\theta} } + \frac{\partial \textbf{g}}{\partial \textbf{x}^{*} } \frac{d\textbf{x}^{*}}{d \boldsymbol{\theta} }
\end{equation}

where $d\textbf{x}^{*}/d\boldsymbol{\theta}$ is evaluated through the implicit function theorem: 

\begin{equation}\label{implicit function theorum}
\frac{d\textbf{x}^{*}}{d \boldsymbol{\theta} } = \left( I-\frac{\partial \textbf{f}_{cycle}}{\partial \textbf{x}^{*} } \right) ^{-1} \frac{\partial \textbf{f}_{cycle}}{\partial  \boldsymbol{\theta} }
\end{equation}

Figure \ref{differentiable model framework} illustrates how these gradients are applied in the differentiable model framework developed in this work. We note that the implicit nature of the CSS criterion implies that the optimiser does not 'see' this constraint. As a result, any error associated with the CSS tolerance can lead to noise in the gradients. In this work, a CSS constraint tolerance of $10^{-5}$ has been used throughout. However, for systems where the spectral radius approaches 1 and tighter optimality tolerances are used within the optimiser, a tighter CSS tolerance might be required.  

\section{Pressure Vacuum Swing Adsorption Case Study}\label{methods}

\begin{figure}
    \centering
    \includegraphics[width=0.75\linewidth]{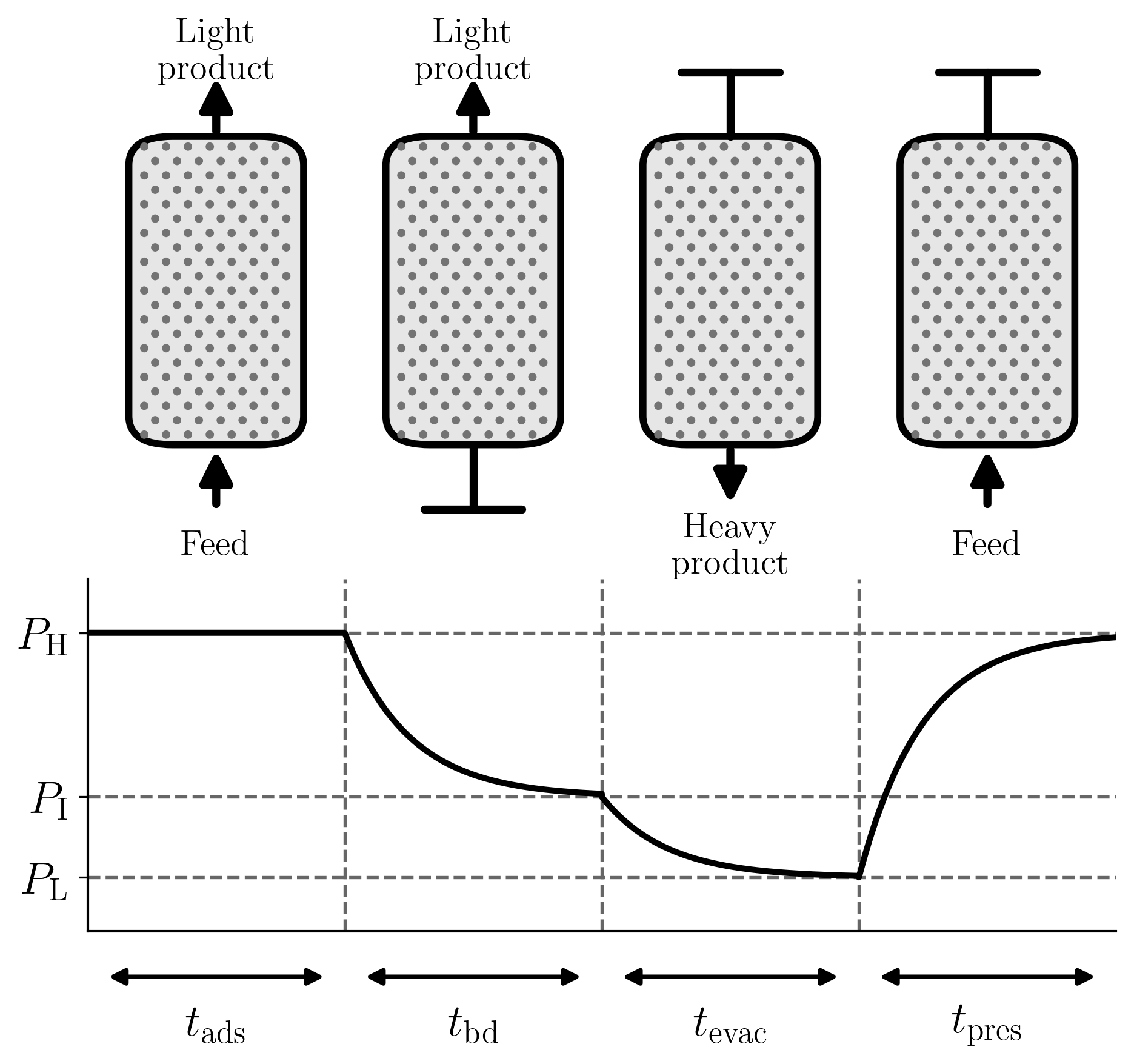}
    \caption{Four-step PVSA cycle. Each step from left to right is: Adsorption (ads), Blowdown (bd), Reverse Evacuation (evac), Feed Pressurisation (pres). In our case study, the feed is a mixture of CO$_\mathrm{2}$ and N$_\mathrm{2}$, the light product is the N$_\mathrm{2}$-rich stream and the heavy product is the CO$_\mathrm{2}$-rich stream. $P_\mathrm{H}$, $P_\mathrm{I}$ and $P_\mathrm{L}$ refer to the high, intermediate and low pressure. The duration of each step is indicated by the horizontal arrows.}
    \label{PVSA graphic}
\end{figure}

The case study considered in this work is a four-step pressure vacuum swing adsorption (PVSA) process for the separation of CO$_\mathrm{2}$ from a dry flue gas containing 15 mol\% CO$_\mathrm{2}$ in N$_\mathrm{2}$ (number of species, $n_\mathrm{c}=2$) at atmospheric pressure and 298~K, using Zeolite 13X as the adsorbent. This system has been the subject of numerous optimisation studies owing to its relevance to post-combustion carbon capture and is therefore a good benchmark for this work.\cite{haghpanah_multiobjective_2013, ward_efficient_2022, subraveti_machine_2019} The four-step cycle consists of (Figure \ref{PVSA graphic}): adsorption at high pressure during which the feed is introduced at one end of the column and the light product (N$_\mathrm{2}$) exits at the other; co-current blowdown to an intermediate pressure which removes most of the residual N$_\mathrm{2}$ from the column; reverse evacuation to the low vacuum pressure during which the heavy product (CO$_2$) is collected; and feed pressurisation back to the adsorption pressure.

A one-dimensional, non-isothermal, non-isobaric model describes the coupled heat and mass transfer within the packed adsorption column. The model comprises partial differential equations for the component mass balances, the overall mass balance, the column internal energy balance, the column wall energy balance and the steady-state momentum balance in the form of Darcy's law for the pressure drop along the column. The following assumptions are adopted in deriving the governing equations:

\begin{itemize}
    \item Only axial gradients are considered (radial gradients within the bed are neglected).
    \item The gas phase obeys the ideal gas law.
    \item Local thermal equilibrium exists between the gas, adsorbed and solid phases within 
    the bed.
    \item The heat capacities of the gas, adsorbed and solid phases are constant.
    \item The densities of the adsorbent and the column wall are constant.
    \item The column void fraction and adsorbent particle radius are uniform and constant.
    \item The axial mass dispersion coefficient and effective thermal conductivity are constant.
    \item The pressure drop along the column is described by Darcy's law.
\end{itemize}

The governing equations are summarised in Table \ref{pde equations}. The state variables are the total pressure $p$, the gas-phase mole fractions $y_i$ of each component $i$ apart from the last, the gas temperature $T$, the column wall temperature $T_\mathrm{w}$, and the solid-phase loadings $q_i$ for all components. In this work, the PDEs are discretised with 10 finite volumes. These equations are accompanied by the adsorption isotherms which describe the thermodynamic equilibrium between the gas and the adsorbed phase. For the Zeolite 13X material used in this work, the extended dual-site Langmuir model is used to quantify this relationship for the CO$_\mathrm{2}$/N$_\mathrm{2}$ mixture:

\begin{equation}\label{DSL equation}
    q_{i}^{*} = \frac{q_{\mathrm{b},i}b_{i}c_{i}}{ 1 + \sum_{i}^{n_\mathrm{c}}b_{i}c_{i}} + \frac{q_{\mathrm{d},i}d_{i}c_{i}}{ 1 + \sum_{i}^{n_\mathrm{c}}d_{i}c_{i}}
\end{equation}

where $q_{\mathrm{b},i}$ and $q_{\mathrm{d},i}$ are the saturation capacities at each adsorption site for each component, $c_{i}=y_ip/RT$ is the concentration of each component in the gas phase, and $b_{i}$ and $d_{i}$ are the equilibrium constants:

\begin{equation}\label{b equilibrium constant}
    b_{i} = b_{0,i} e^{\frac {-\Delta U_{\mathrm{b},i}} {RT} }
\end{equation}

\begin{equation}\label{d equilibrium constant}
    d_{i} = d_{0,i} e^{\frac {-\Delta U_{\mathrm{d},i}} {RT} }
\end{equation}

The isotherm parameters are taken from the unequal energy sites fitting from \citeauthor{wilkins_measurement_2019}.\cite{wilkins_measurement_2019} The mass transfer resistance for the diffusion from the bulk gas to the adsorption site is described through the linear driving force (solid phase material balance in Table \ref{pde equations}) with a constant mass transfer coefficient, $k$. Lastly, the longitudinal dispersion coefficient is found through the following correlation.\cite{ruthven_principles_1984}

\begin{equation}\label{Dispersion correlation}
    D_{\mathrm{L}} = 0.7D_{\mathrm{m}} + v_{\mathrm{feed}}r_{\mathrm{p}}
\end{equation}

where $D_\mathrm{m}$ is a molecular diffusion coefficient and $v_\mathrm{feed}$ is the superficial feed velocity. Values of the parameters used within this model are given within Table \ref{parameters}. The boundary conditions for each step are set according to the direction of flow and the type of operation. During the adsorption step, the feed-end boundary imposes Danckwerts conditions on the mole fractions and temperature, and a fixed inlet velocity. During blowdown and evacuation steps, an exponential pressure profile is imposed at the product end of the column, whilst zero-flux conditions are applied at the closed end. During the pressurisation step, an exponential pressure profile is prescribed at the feed boundary. The boundary conditions for each step are described in Section S5 of the Supporting Information. 

\begin{table}
\centering
\renewcommand{\arraystretch}{2.5}
\begin{tabular}{l}
\hline
\textbf{Overall material balance} \\
$\dfrac{\partial p}{\partial t} - \dfrac{p}{T}\dfrac{\partial T}{\partial t}
= -T \dfrac{\partial}{\partial z}\!\left(\dfrac{pv}{T}\right)
- RT\dfrac{\rho_{\mathrm{b}}}{\epsilon}\sum_{i=1}^{n_\mathrm{c}} \dfrac{\partial q_i}{\partial t}$ \\[6pt]
\hline
\textbf{Component material balance for species $i \in \{1, \ldots, n_\mathrm{c}-1\}$} \\
$\dfrac{\partial y_i}{\partial t}
= \dfrac{T}{p}D_{\mathrm{L}}\dfrac{\partial}{\partial z}\!\left(\dfrac{p}{T}\dfrac{\partial y_i}{\partial z}\right)
- \dfrac{T}{p}\left(\dfrac{\partial}{\partial z}\!\left(\dfrac{y_i p v}{T}\right)
- y_i \dfrac{\partial}{\partial z}\!\left(\dfrac{pv}{T}\right)\right)
- \dfrac{RT}{p}\dfrac{\rho_{\mathrm{b}}}{\epsilon}
  \left(y_i \sum_{i=1}^{n_\mathrm{c}}\dfrac{\partial q_i}{\partial t}
  - \dfrac{\partial q_i}{\partial t}\right)$ \\[6pt]
\hline
\textbf{Solid phase material balance for species $i \in \{1, \ldots, n_\mathrm{c}\}$} \\
$\dfrac{\partial q_i}{\partial t} = k\!\left(q_i^* - q_i\right)$ \\[6pt]
\hline
\textbf{Column energy balance} \\
$\left[C_{\mathrm{p,g}}\dfrac{P}{RT}
+ \dfrac{\rho_{\mathrm{b}}}{\epsilon}\!\left(C_{\mathrm{p,s}}
+ C_{\mathrm{p,a}}\sum_{i=1}^{n_\mathrm{c}} q_i\right)\right]\dfrac{\partial T}{\partial t}
= \dfrac{K_{\mathrm{z}}}{\epsilon}\dfrac{\partial^2 T}{\partial z^2}
- \dfrac{C_{\mathrm{p,g}}}{R}\!\left(\dfrac{\partial(pv)}{\partial z}
- T\dfrac{\partial}{\partial z}\!\left(\dfrac{pv}{T}\right)\right)$ \\
$\hspace{4cm}- \dfrac{\rho_{\mathrm{b}}}{\epsilon}T\!\left(C_{\mathrm{p,a}}
- C_{\mathrm{p,g}}\right)\sum_{i=1}^{n_\mathrm{c}}\dfrac{\partial q_i}{\partial t}
+ \dfrac{\rho_{\mathrm{b}}}{\epsilon}\sum_{i=1}^{n_\mathrm{c}}\!\left(-\Delta H_i
\dfrac{\partial q_i}{\partial t}\right)
- \dfrac{2h_{\mathrm{in}}}{\epsilon r_{\mathrm{in}}}\!\left(T - T_{\mathrm{w}}\right)$ \\[6pt]
\hline
\textbf{Column wall energy balance} \\
$\rho_{\mathrm{w}} C_{\mathrm{p,w}}\dfrac{\partial T_{\mathrm{w}}}{\partial t}
= K_{\mathrm{w}}\dfrac{\partial^2 T_{\mathrm{w}}}{\partial z^2}
+ \dfrac{2r_{\mathrm{in}} h_{\mathrm{in}}}{r_{\mathrm{out}}^2 - r_{\mathrm{in}}^2}\!\left(T - T_{\mathrm{w}}\right)
- \dfrac{2r_{\mathrm{out}} h_{\mathrm{out}}}{r_{\mathrm{out}}^2 - r_{\mathrm{in}}^2}\!\left(T_{\mathrm{w}} - T_{\mathrm{a}}\right)$ \\[6pt]
\hline
\textbf{Steady-state momentum balance (Ergun equation)} \\
$-\dfrac{\partial p}{\partial z}
= \dfrac{150}{4r_{\mathrm{p}}^2}\!\left(\dfrac{1-\epsilon}{\epsilon}\right)^{2}\mu v$ \\[6pt]
\hline
\end{tabular}
\caption{Mass, energy and momentum balances for the 1-dimensional axially dispersed plug flow model}
\label{pde equations}
\end{table}

\subsection{Isosteric heat of adsorption}
\label{isosteric heat}

The isosteric heat of adsorption describes the enthalpy released during adsorption. The Clausius-Clapeyron equation for the isosteric heat of adsorption for a multicomponent ideal gas mixture in equilibrium with an adsorbent is given as:\cite{sircar_estimation_1992}

\begin{equation}
  -\Delta H_{\mathrm{iso},i} = R T^{2} \left( \frac{\mathrm{d} \ln\!\left( P\, y_{i} \right)}{\mathrm{d} T} \right)_{ \mathbf{q} } = R T^{2} \left[ \frac{1}{P}\!\left(\frac{\mathrm{d} P}{\mathrm{d} T}\right)_{\!\mathbf{q}} + \frac{1}{y_{i}}\!\left(\frac{\mathrm{d} y_{i}}{\mathrm{d} T}\right)_{\!\mathbf{q}} \right]
  \label{clausius clapeyron}
\end{equation}

The subscript $\mathbf{q}=(q_{1},\dots,q_{\mathrm{nc}})$ indicates that the derivative is evaluated at constant adsorbed-phase loading of every component. As the DSL isotherm describes loading as a function of the gas phase pressure, temperature and composition, we apply implicit differentiation to this isotherm using AD in JAX. At constant loading ($dq_i=0$), the following expression is derived for all components: 

\begin{equation}
  0 \;=\; \frac{\partial q_{i}}{\partial P}\, \frac{\mathrm{d} P}{\mathrm{d} T}
  \;+\; \frac{\partial q_{i}}{\partial T}
  \;+\; \sum_{j=1}^{n_\mathrm{c}-1}
        \frac{\partial q_{i}}{\partial y_{j}}\, \frac{\mathrm{d} y_{j}}{\mathrm{d} T},
  \label{eq:implicit}
\end{equation}

Equation~\ref{eq:implicit} is solved simultaneously for each component $i=1..n_\mathrm{c}$, whereby $(\mathrm{d}y_{n_\mathrm{c}}/\mathrm{d}T)_{\mathbf{q}} =
 -\sum_{j=1}^{n_{c}-1}(\mathrm{d}y_{j}/\mathrm{d}T)_{\mathbf{q}}$. The derivatives of pressure, and all mole fractions with respect to temperature at constant loading are used in Equation~\ref{clausius clapeyron} to estimate the isosteric heat of adsorption. 

\begin{table}[h]
\footnotesize
    \centering
    \begin{tabular}{l c c}
        \hline
        Column length, $L$ & [m] & 1.0 \\ 
        Column inner radius, $r_\mathrm{in} $ & [m] & 0.145 \\
        Column outer radius, $r_\mathrm{out} $ & [m] & 0.162 \\
        Bed density, $\rho_\mathrm{b}$ & [kg/m\(^3\)] & 712 \\
        Bed voidage, $\epsilon$ & & 0.37 \\
        Particle voidage, $\epsilon_\mathrm{p}$ & & 0.35 \\
        Particle radius, $r_\mathrm{p}$ & [m] & 0.001 \\
        Particle tortuosity, $\tau_\mathrm{p}$ & & 3 \\
        Molecular diffusivity, $D_\mathrm{m}$ & [m\(^2\)/s] & 1.5$\times$ 10\(^{-7}\) \\
        Thermal conductivity of gas, $K_\mathrm{z}$ & [J/m/K/s] & 0.09 \\
        Thermal conductivity of wall, $K_\mathrm{w}$ & [J/m/K/s] & 16 \\
        Heat capacity of gas phase, $C_\mathrm{pg}$ & [J/mol/K] & 30.7 \\
        Heat capacity of adsorbed phase, $C_\mathrm{pa}$ & [J/mol/K] & 30.7 \\
        Heat capacity of adsorbent, $C_\mathrm{ps}$ & [J/kg/K] & 1,070 \\
        Heat capacity of column wall, $C_\mathrm{pw}$ & [J/kg/K] & 502  \\
        Density of wall, $\rho_\mathrm{w}$ & [kg/m\(^3\)] & 7,800  \\
        Dynamic viscosity of gas, $\mu$ & [kg/m/s] & 1.72 $\times$ 10\(^{-5}\) \\
        Overall inside heat transfer coefficient, $h_\text{in}$ & [J/m\(^2\)/K/s] & 8.6 \\
        Overall outside heat transfer coefficient, $h_\text{out}$ & [J/m\(^2\)/K/s] & 2.5 \\
        Pressure profile time constant, $\lambda$ & [/s] & 0.5 \\
        Compressor efficiency, $\eta$ & & 0.72 \\
        Mass transfer coefficient (LDF), $k$ & [/s] & 26.25 \\
        Saturation capacity of CO$_\mathrm{2}$ at site 1, $q_\mathrm{b,CO_{2}}$ & [mol/kg] & 3.257 \\
        Saturation capacity of CO$_\mathrm{2}$ at site 2, $q_\mathrm{d,CO_{2}}$ & [mol/kg] & 3.24 \\
        Saturation capacity of N$_\mathrm{2}$ at site 1, $q_\mathrm{b,N_{2}}$ & [mol/kg] & 3.257 \\
        Saturation capacity of N$_\mathrm{2}$ at site 2, $q_\mathrm{d,N_{2}}$ & [mol/kg] & 3.24 \\
        Adsorption equilibrium constant of CO$_\mathrm{2}$ at site 1, $b_\mathrm{CO_{2},0}$ & [m\(^3\)/mol] & 2.09 $\times$ 10\(^{-7}\) \\
        Adsorption equilibrium constant of CO$_\mathrm{2}$ at site 2, $d_\mathrm{CO_{2},0}$ & [m\(^3\)/mol] & 1.06 $\times$ 10\(^{-7}\) \\
        Adsorption equilibrium constant of N$_\mathrm{2}$ at site 1, $b_\mathrm{N_{2},0}$ & [m\(^3\)/mol] & 7.96 $\times$ 10\(^{-7}\) \\
        Adsorption equilibrium constant of N$_\mathrm{2}$ at site 2, $d_\mathrm{N_{2},0}$ & [m\(^3\)/mol] & 6.94 $\times$ 10\(^{-7}\) \\
        Molar internal energy change for adsorption of CO$_\mathrm{2}$ at site 1, $-\Delta U_\mathrm{b,CO_{2}}$ & [kJ/mol] & 42.67 \\
        Molar internal energy change for adsorption of CO$_\mathrm{2}$ at site 2, $-\Delta U_\mathrm{d,CO_{2}}$ & [kJ/mol] & 32.21 \\
        Molar internal energy change for adsorption of N$_\mathrm{2}$ at site 1, $-\Delta U_\mathrm{b,N_{2}}$ & [kJ/mol] & 18.86 \\
        Molar internal energy change for adsorption of N$_\mathrm{2}$ at site 2, $-\Delta U_\mathrm{d,N_{2}}$ & [kJ/mol] & 17.76 \\
        
    \end{tabular}
    \caption{Parameters used within adsorption model simulation}
    \label{parameters}
\end{table}

\subsection{Key Performance Indicators}

The four KPIs that are used for the design of the PVSA process include the purity of the CO$_\mathrm{2}$ product ($Pu_{\mathrm{CO_{2}}}$), the recovery of CO$_\mathrm{2}$ ($Re_{\mathrm{CO_{2}}}$), the productivity of the process ($Pr$) and the energy consumption associated with the compressors and vacuum pumps ($E_\mathrm{T}$). For the calculation of these KPIs, three further process units are integrated in time along with the states of the column. These include an isentropic compressor, an isentropic vacuum pump and a tank to calculate the quantity and composition of gas discharged from the outlet of the adsorption step. The power consumption of the compressor and vacuum pump as well as the total gas and CO$_\mathrm{2}$ molar flowrate from the outlet of the adsorption step are added as state variables for the ODE solution. The equations for these units and the KPIs are given in Sections S7 and S8 of the Supporting Information. 

\subsection{Process optimisation}

The design of adsorption processes for post combustion carbon capture applications typically involve multi-objective optimisation of two conflicting KPIs. The first involves the maximisation of the purity and recovery, which we formulate as a constrained single-objective optimisation problem: 

\begin{equation}\label{unconstrained optimisation problem}
    \begin{aligned}
    \min_{\boldsymbol{\theta} } \quad & - Re_\mathrm{CO_{2}}(\boldsymbol{\theta})\\
    \textrm{s.t.} \quad & Pu_\mathrm{CO_{2}}(\boldsymbol{\theta}) \ge \varepsilon\\
      & \boldsymbol{\theta}\in[\boldsymbol{\theta}^\mathrm{L},\boldsymbol{\theta}^\mathrm{U}]    \\
    \end{aligned}
\end{equation}

where $\varepsilon$ is suitably varied so as to identify the Pareto curve of optimal solutions (epsilon-constraint method\cite{miettinen_nonlinear_1998}). The second optimisation problem uses an analogous formulation to maximise productivity and minimise energy consumption, albeit under constraints on purity and recovery: 

\begin{equation}\label{constrained optimisation problem}
    \begin{aligned}
    \min_{\boldsymbol{\theta} } \quad & E_\mathrm{T}(\boldsymbol{\theta})\\
    \textrm{s.t.} \quad & Pr(\boldsymbol{\theta}) \ge \varepsilon \\
    & Pu_\mathrm{CO_{2}}(\boldsymbol{\theta}) \ge 95\% \\
    & Re_\mathrm{CO_{2}}(\boldsymbol{\theta}) \ge 90\% \\
      & \boldsymbol{\theta}\in[\boldsymbol{\theta}^\mathrm{L},\boldsymbol{\theta}^\mathrm{U}]    \\
    \end{aligned}
\end{equation}

For these optimisation problems, the design variables are bounded with the values given in Table \ref{design variable bounds}, and include the duration and pressure of the adsorption step ($t_\mathrm{ads}$ and $P_H$, respectively), the feed velocity ($v_{\mathrm{feed}}$), the duration  and pressure of the blowdown ($t_{\mathrm{bd}}$ and $P_{\mathrm{I}}$, respectively), and the duration of the evacuation step ($t_{\mathrm{evac}}$). For the design of the PVSA process, the low pressure set point was fixed at 0.05~bar and the pressurisation time at 20~s. 

\begin{table}[h]
\centering
\begin{tabular}{lcccccc}
\toprule
\textbf{Variable} 
    & $t_{\mathrm{ads}}$ 
    & $P_{\mathrm{H}}$ 
    & $v_{\mathrm{feed}}$ 
    & $t_{\mathrm{bd}}$ 
    & $P_{\mathrm{I}}$ 
    & $t_{\mathrm{evac}}$ \\
\midrule
\textbf{Unit} 
    & s 
    & bar 
    & m\,s$^{-1}$ 
    & s 
    & bar 
    & s \\[4pt]
\textbf{Lower Bound} 
    & 20 & 1 & 0.1 & 30 & 0.07 & 30 \\
\textbf{Upper Bound} 
    & 100 & 10 & 2 & 100 & 3 & 100 \\
\bottomrule
\end{tabular}
\caption{Design variable bounds for the optimisation of the 4-step PVSA process}
\label{design variable bounds}
\end{table}

The two optimisation problems described above are solved using the IPOPT nonlinear programming algorithm.\cite{wachter_implementation_2006} Although IPOPT is employed here, the overall optimisation framework is readily transferable to other gradient-based optimisers and thus the methodology is not restricted to a single numerical solver implementation. Furthermore, due to the nonconvexity of the optimisation problem, this gradient-based algorithm could converge to local minima. To explore this, a multi-start is conducted for each optimisation problem, where ten runs of the algorithm are completed, each from a random start point within the design space. The random point is not required to satisfy the constraints. The gradient-based approach is compared to the NSGA-II algorithm, which is widely used in the adsorption literature for the multi-objective black-box optimisation, where a population size of 72 with 70 generations was used. Sobol sampling of the design space is also used for this comparison with 4096 samples. All the simulations were performed on an Intel Core i5-10500 CPU with 8GB of RAM.

\section{Results}\label{results}

The performance of the end-to-end differentiable model is quantified at various levels. First, we compare the simulation results with those from a similar model within the literature. Second, we compare by means of forward simulations the direct determination method to accelerate the convergence to CSS to the successive substitution method. Lastly, we compare results obtained by solving various multi-objective optimisation problems. 

\subsection{Simulation verification}

The framework developed in this work is based upon the standard model formulation used throughout the adsorption literature.\cite{haghpanah_multiobjective_2013, ward_efficient_2022, wilkins_pilot-scale_2025, pai_superstructure_2025, peh_metal-organic_2022, subraveti_vector_2026, subraveti_machine_2019} Before the methods introduced in this work are compared, the model is verified against published results to confirm that the implementation produces consistent predictions. The minimum energy designs reported by \citeauthor{haghpanah_multiobjective_2013}\cite{haghpanah_multiobjective_2013} are used for this comparison, where constraints of 90\% purity and 90\% recovery are enforced. The operating conditions for each design are taken directly from the reported optimal solutions and evaluated in the present model without re-optimisation. The model parameters used for this comparison are provided in Section S9 of the Supporting Information.
 
Table \ref{tab:model_verification} compares the productivity and specific energy consumption obtained from the present model with the values reported by \citeauthor{haghpanah_multiobjective_2013} across five minimum energy designs at different evacuation pressures. The two models show good agreement, with deviations in productivity within $\pm$5\% and deviations in energy consumption within 4\%. The purity and recovery values obtained from the present model are consistent with the enforced constraints. Minor differences between the two implementations exist in the model formulation and in the calculation of the key performance indicators (Section S9 of the Supporting Information).
 
\begin{table}[h]
    \centering
    \small
    \setlength{\tabcolsep}{4pt}
    \caption{Comparison of the present model with the results of \citeauthor{haghpanah_multiobjective_2013}\cite{haghpanah_multiobjective_2013} for the minimum energy designs at different evacuation pressures. Productivity is given in mol\,m$^{-3}$\,s$^{-1}$ and energy consumption in kWh\,t$^{-1}$. $\Delta$ denotes the relative difference between the present model and the reference values.}
    \label{tab:model_verification}
    \begin{tabular}{ccccccccc}
        \toprule
        & & & \multicolumn{3}{c}{\textbf{Productivity}} & \multicolumn{3}{c}{\textbf{Energy consumption}} \\
        \cmidrule(lr){4-6} \cmidrule(lr){7-9}
        $\boldsymbol{P_L}$ \textbf{[bar]} & \textbf{Purity} & \textbf{Recovery} & \textbf{This work} & \textbf{Ref.} & $\boldsymbol{\Delta}$ \textbf{[\%]} & \textbf{This work} & \textbf{Ref.} & $\boldsymbol{\Delta}$ \textbf{[\%]} \\
        \midrule
        0.02 & 0.906 & 0.894 & 0.297 & 0.312 & $-$4.97 & 143.6 & 149.0 & $-$3.62 \\
        0.03 & 0.906 & 0.907 & 0.508 & 0.512 & $-$0.87 & 194.5 & 201.5 & $-$3.46 \\
        0.05 & 0.903 & 0.909 & 0.333 & 0.348 & $-$4.42 & 289.3 & 298.9 & $-$3.23 \\
        0.07 & 0.910 & 0.914 & 0.704 & 0.688 & $+$2.35 & 373.2 & 388.7 & $-$3.98 \\
        0.10 & 0.918 & 0.911 & 0.721 & 0.694 & $+$3.86 & 486.8 & 499.7 & $-$2.56 \\
        \bottomrule
    \end{tabular}
\end{table}

\subsection{Acceleration to CSS}\label{Acceleration to CSS results}

\begin{figure}
    \centering
    \includegraphics[width=1\linewidth]{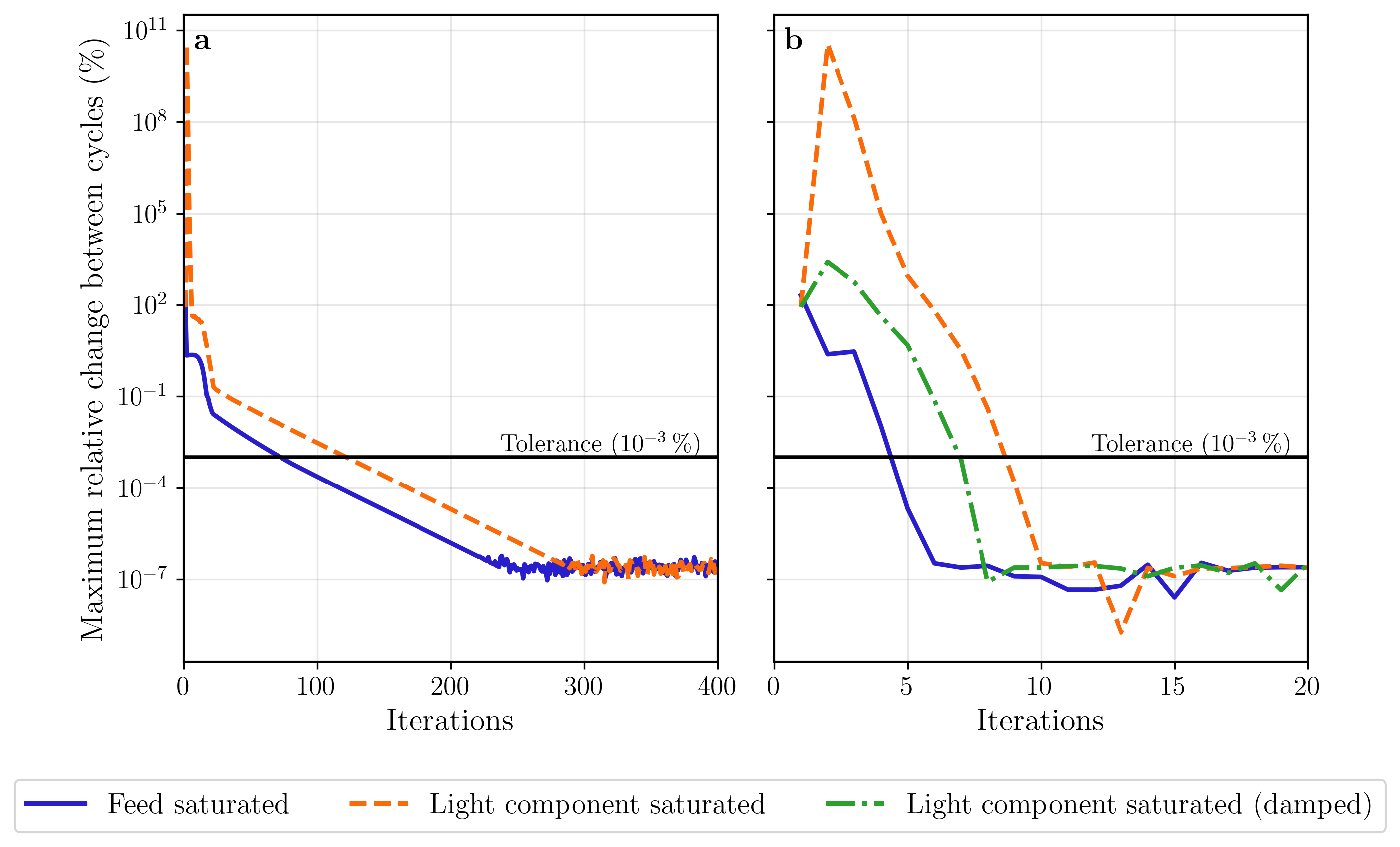}
    \caption{Convergence plots for the successive substitution (left) and direct determination (right) methods for two initial conditions: feed- (solid line) and nitrogen-saturated bed (dashed line). In the direct determination method, two different fallback methods were used when negative mole fractions or adsorbed quantities were encountered successive substitution (dashed line) and damped Newton (dotted-dashed line). Design parameters: $P_\mathrm{H}=8$~bar, $P_\mathrm{I}=1.5$~bar, $v_\mathrm{feed}=0.8$~ms$^{\mathrm{-1}}$, $t_\mathrm{ads}=50$~s, $t_\mathrm{bd}=100$~s, $t_\mathrm{evac}=100$~s.}
    \label{convergence plots}
\end{figure}

The convergence performance to CSS of the successive substitution and direct determination methods is compared for two different initial conditions in Figure \ref{convergence plots}: a bed saturated with the feed gas and a bed saturated with N$_\mathrm{2}$ (the light component). The successive substitution method (Figure \ref{convergence plots}a) requires 82 and 115 iterations to reach the CSS tolerance from a bed saturated with feed and with N$_\mathrm{2}$, respectively. In both scenarios, linear convergence is exhibited, as expected for this fixed point iteration, and nearly 300 iterations are required to reach the error associated with the solver tolerance. In contrast, the direct determination method (Figure \ref{convergence plots}b) requires significantly fewer iterations to attain CSS (between 5 and 10 iterations, depending on the scenario), reflecting the near-quadratic convergence behaviour expected for a Newton iteration near the solution. The scenario with the bed saturated with the feed gas as the initial condition features the best convergence, with just 5 iterations required to reach CSS. The scenarios with the bed saturated with N$_\mathrm{2}$ show a slightly lower performance, because either the damping fallback method or the successive substitution method had to be deployed as fall-back options when the (undampened) Newton iteration predicted negative values of mole fraction and adsorbed quantity for the next iteration. A similar observation was reported by \citeauthor{croft_periodic_1994}\cite{croft_periodic_1994} in their air purification example where the initial state is an empty bed. In our study, the damping fallback method provides an improved convergence to the CSS than reverting to the successive substitution method. 

\begin{figure}
    \centering
    \includegraphics[width=0.6\linewidth]{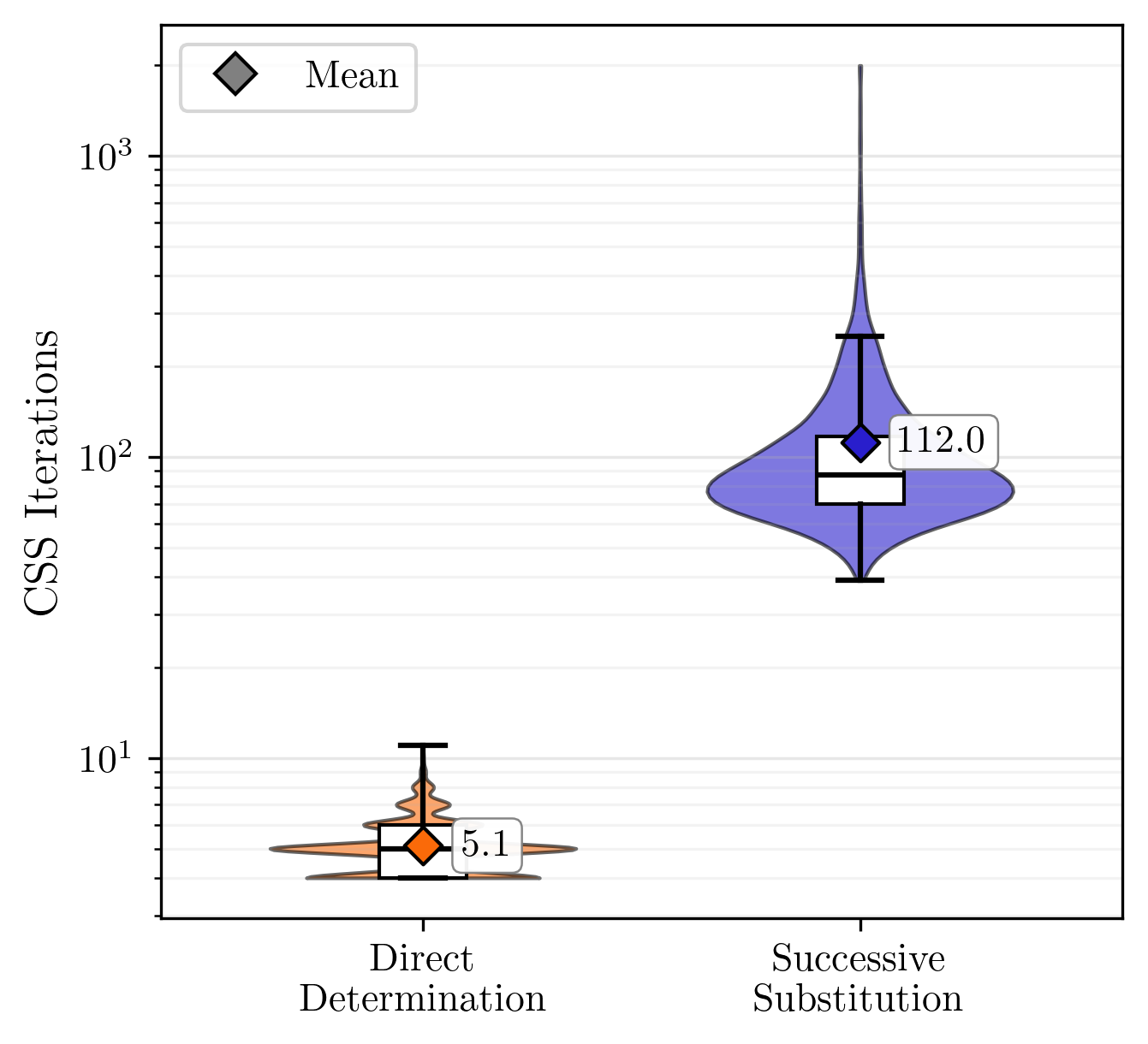}
    \caption{Violin plots of the number of iterations required to reach CSS for the direct determination and successive substitution methods across 4096 Sobol sampled designs. The white box illustrates the quartiles of the distribution and the whiskers correspond to 1.5 times the interquartile range. The mean of the distribution is indicated by the diamond symbol.}
    \label{violin plots iterations}
\end{figure}

The two methods are again compared in Figure \ref{violin plots iterations} by considering 4096 different designs in the six-dimensional design space (Table \ref{design variable bounds}). To this end, we have considered a cycle that begins with the high pressure adsorption step (the bed is saturated with the feed gas). In the figure, the violin plots show the number of iterations required to reach CSS for all 4096 samples and for the two methods (the sample set was identical for both methods). The direct determination method requires on average just 5.1 iterations to reach CSS with all of the 4096 designs converging within 11 iterations. The successive substitution method requires on average 112 cycles to reach CSS, with a much wider distribution of number of iterations (40--2000) compared to the direct determination method. 

The mass balance error (Equation 57 in the Supporting Information) of the cycle at CSS, as well as the difference between the KPIs obtained for the two methods are given in Table \ref{SS vs DD results table}. Both methods feature negligibly small errors in the mass balance, with maximum errors of 0.003\% and 0.0001\% for the successive substitution and the direct determination method, respectively. Across all 4096 designs, there is excellent agreement of the KPIs computed through both methods, with all designs being within 0.1\% of each other for the four KPIs recorded. This shows that the two methods converge to the same CSS for all the designs explored. 

It is possible that a process cycle can exhibit multiple periodic states or have a periodic state that repeats every 2 cycles, but this was not observed for this process.\cite{croft_periodic_2} Furthermore, it is possible the direct determination method could converge to a point that is mathematically a CSS, but physically unattainable. A CSS of this type is unstable such that any disturbance would cause the system to diverge from this CSS. Mathematically, this CSS would have a spectral radius greater than 1. The spectral radius for the design corresponding to the convergence plots in Figure \ref{convergence plots} is 0.9509, confirming a stable and physically plausible CSS. The value of this spectral radius also gives an insight into the cycle dynamics where values closer to unity exhibit slower cycle dynamics and would represent a process that requires more cycles to reach CSS. 

\begin{table}[h]
\centering
\begin{threeparttable}
\begin{tabular}{l
                S[table-format=1.2e-1]
                S[table-format=1.2e-1]
                S[table-format=1.2e-1]
                S[table-format=1.2e-1]}
\toprule
& \multicolumn{2}{c}{\textbf{Successive Substitution}}
& \multicolumn{2}{c}{\textbf{Direct Determination}} \\
\cmidrule(lr){2-3} \cmidrule(lr){4-5}
& \textbf{Mean (\%)} & \textbf{Max (\%)}
& \textbf{Mean (\%)} & \textbf{Max (\%)} \\
\midrule
\multicolumn{5}{l}{\textit{Mass Balance Error}} \\
\cmidrule(l){1-5}
\quad Overall 
    & 1.86e-04  & 3.15e-03
    & 4.26e-06  & 1.30e-04 \\
\midrule
\multicolumn{5}{l}{\textit{KPI Difference vs.\ Successive Substitution}\tnote{a}} \\
\cmidrule(l){1-5}
\quad Purity          
    & {---} & {---} & 9.31e-04 & 2.51e-02 \\
\quad Recovery        
    & {---} & {---} & 7.83e-04 & 6.99e-03 \\
\quad Productivity    
    & {---} & {---} & 1.36e-03 & 5.59e-02 \\
\quad Energy consumption 
    & {---} & {---} & 7.94e-04 & 3.96e-03 \\
\bottomrule
\end{tabular}
\begin{tablenotes}
    \footnotesize
    \item[a] KPI differences are computed as the absolute percentage deviation 
             of the Direct Determination method relative to Successive 
             Substitution as the reference solution.
\end{tablenotes}
\end{threeparttable}
\caption{Mass balance errors and KPI differences between the successive 
         substitution and direct determination methods.}
\label{SS vs DD results table}
\end{table}

\subsection{PVSA process optimisation}

\subsubsection{Three decision variables}

The direct determination method provides acceleration to the solution speed of each individual simulation for use within an optimisation framework. For this design problem, three different approaches are compared: Sobol sampling of the design space (4096 samples), the NSGA-II algorithm and the gradient-based optimisation using the IPOPT algorithm. The latter exploits the gradient information readily available from the differentiable model. The comparison is first performed for a reduced design problem involving three decision variables: the gas feed velocity, $v_\mathrm{feed}$, the high pressure set point, $P_\mathrm{H}$, and the intermediate pressure set point, $P_\mathrm{I}$. The adsorption, blowdown and evacuation step durations are fixed at 50, 100 and 100 seconds, respectively \cite{sachio_operability-economics_2024}. 

\begin{figure}[h]
    \centering
    \includegraphics[width=0.75\linewidth]{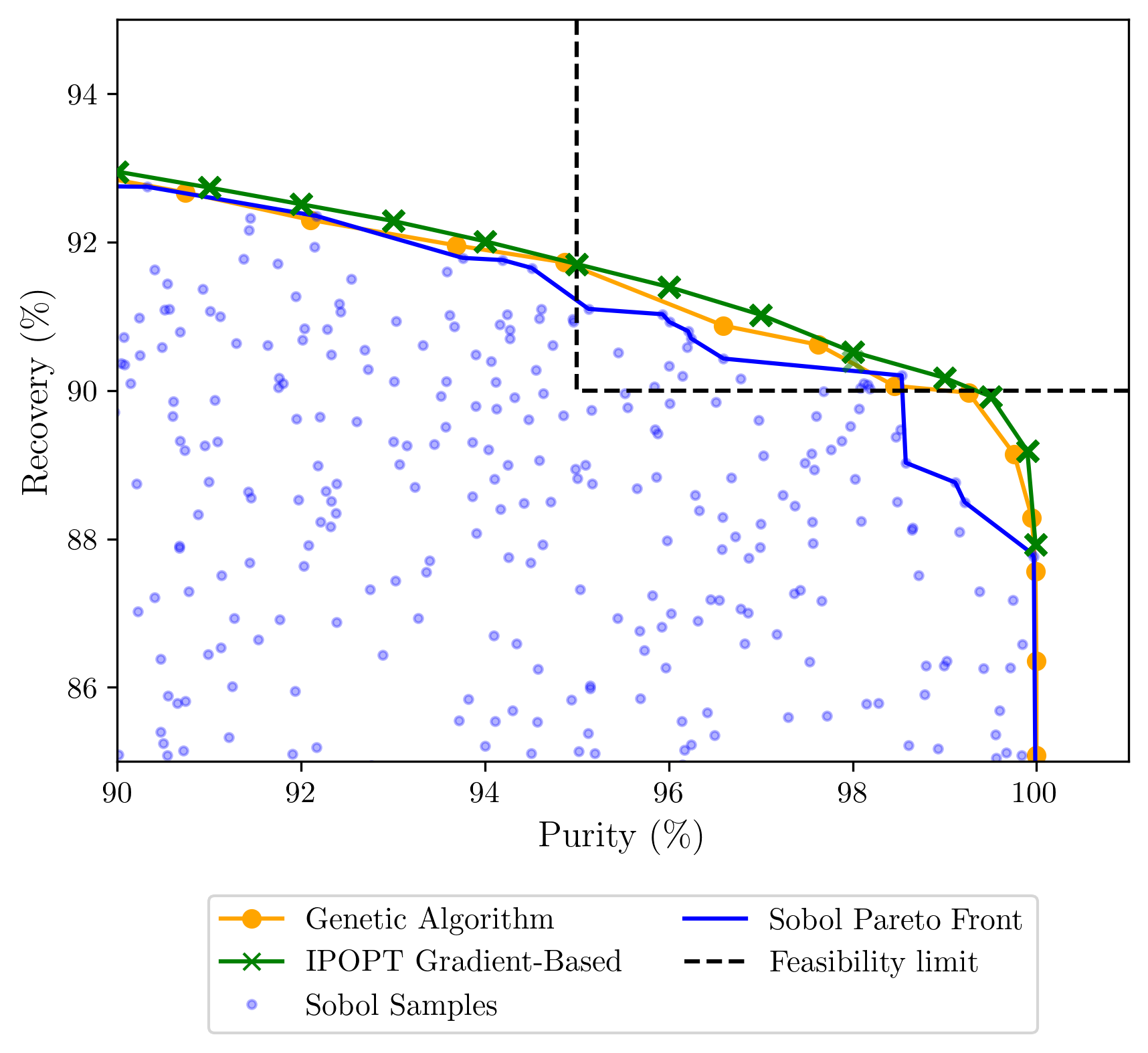}
    \caption{Pareto front for the purity-recovery optimisation problem solved with Sobol sampling, the NSGA-II algorithm and the IPOPT algorithm which uses gradients from the differentiable model. For the latter,  a multi-start strategy (10 random initial start points) is adopted and all runs are shown with the green crosses.}
    \label{3 variable optimisation unconstrained}
\end{figure}

Figure \ref{3 variable optimisation unconstrained} shows the Pareto fronts obtained for the multi-objective optimisation problem where both purity and recovery are maximised (Equation \ref{unconstrained optimisation problem}). To construct the gradient-based Pareto front, the epsilon constraint on purity is varied between 90\% and 99.5\% in 0.5\% increments, with additional evaluations at 99.9\% and 99.99\% purity. The three solution methods identify similar Pareto fronts and capture the trade-off between CO$_2$ purity and CO$_2$ recovery. The gradient-based optimisation yields a slightly improved front than the genetic algorithm, with both methods improving upon Sobol sampling, which produces a cloud of feasible designs lying close to but below the optimal front. The multi-start strategy adopted for the gradient-based optimisation reveals that the non-convexity of this problem is significant in the sub-problem with the 98\% purity constraint. Two of the ten multi-start points converge to a distinct sub-optimal point, characterised by a larger intermediate pressure and feed velocity, while the high pressure converges to its upper bound in all cases. The translucent green crosses in Figure \ref{3 variable optimisation unconstrained} indicate these locally optimal points. The raw data for all gradient-based optimisation runs are provided in the Supplementary Information.

\begin{figure}[h]
    \centering
    \includegraphics[width=0.75\linewidth]{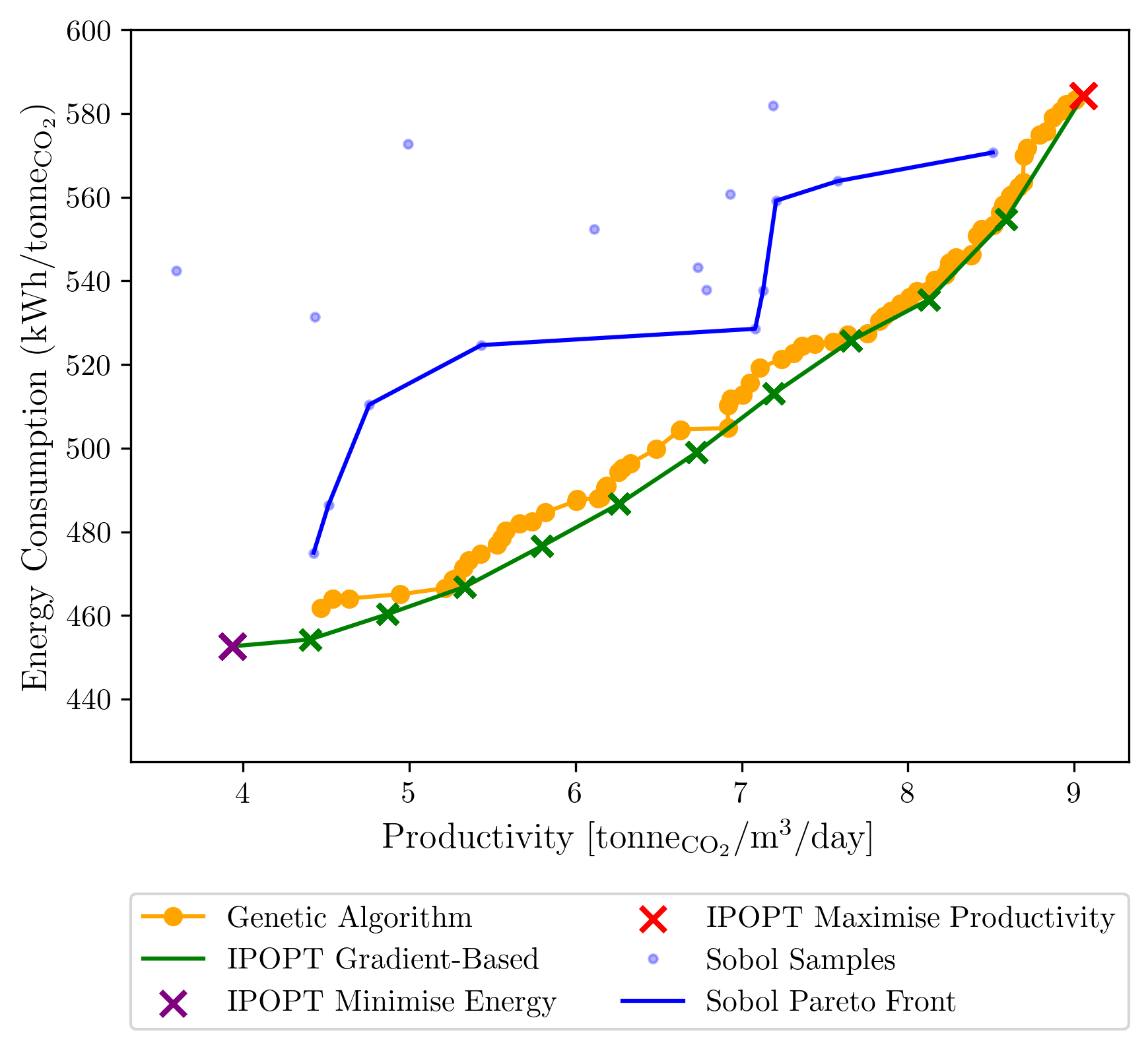}
    \caption{Pareto front for the minimisation of energy consumption and maximisation of productivity with 95\% purity and 90\% recovery constraints comparing Sobol sampling, the NSGA-II algorithm and the IPOPT algorithm which uses gradients from the differentiable model. Scenarios with three decision variables.}
    \label{3 variable optimisation constrained}
\end{figure}

Figure \ref{3 variable optimisation constrained} shows the Pareto fronts obtained for the multi-objective optimisation problem where productivity is maximised and energy usage is minimised (Equation \ref{constrained optimisation problem}) - again, for the three methods. The genetic algorithm and the gradient-based optimisation produce Pareto fronts that are in close agreement, while the approach that uses the 4096 Sobol samples demonstrates obvious deficiencies. In fact, as shown in Figure \ref{3 variable optimisation unconstrained}, relatively few points in the whole design space satisfy  simultaneously the constraints on CO$_2$ purity (95\%) and  CO$_2$ recovery (90\%). In other words, for the same number of Sobol samples, a reduction in the size of the sampled input space would be needed to increase the frequency of feasible designs. In contrast to the purity/recovery optimisation problem, every multi-start point for each epsilon-constrained sub-problem converges to the same solution. The added purity and recovery constraints restrict the feasible region for this optimisation which likely reduces the possibility of converging to multiple optima. The gradient-based algorithm is also used to evaluate single-objective optimisation problems for maximising the productivity (indicated by the red cross) and for minimising the energy consumption (indicated by the purple cross) and to explore the extreme ends of the Pareto front. We note that neither the genetic algortihm, nor the Sobol sampling points reached the point with the lowest energy consumption, highlighting the value of the single-objective gradient-based optimisation to capture the extremes of the Pareto curve. The convergence of the multi-start to identical points in the gradient-based optimisation and the strong agreement with the NSGA-II optimiser provide confidence that the identified Pareto front closely represents the true optimal trade-off for this three-variable problem. 

\begin{figure}[h]
    \centering
    \includegraphics[width=0.75\linewidth]{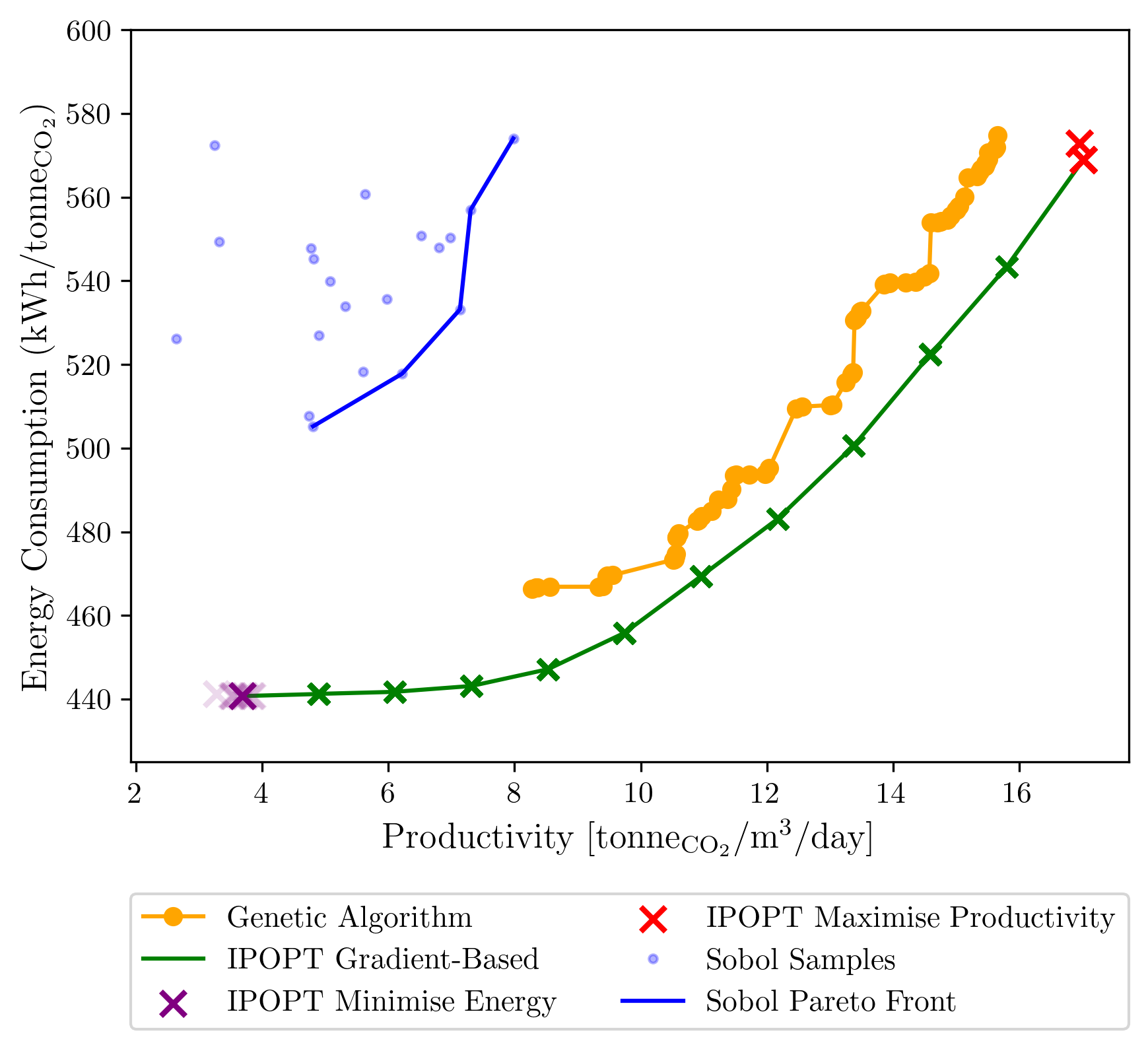}
    \caption{Pareto front for the minimisation of energy consumption and maximisation of productivity with 95\% purity and 90\% recovery constraints comparing Sobol sampling, the NSGA-II algorithm and the IPOPT algorithm which uses gradients from the differentiable model. Scenarios with six decision variables.}
    \label{constrained pareto front 6 variables}
\end{figure}

\subsubsection{Six decision variables}

We explore in this section the full design problem to include all six decision variables in Table~\ref{design variable bounds}. Figure \ref{constrained pareto front 6 variables} shows the Pareto fronts obtained with the same three methods as in the previous section. For this higher-dimensional problem, the gradient-based optimisation produces a Pareto front that is significantly improved over the one produced by the genetic algorithm. Notably, the latter also shows a shorter Pareto curve and doesn't extend into the region of low and high productivity bounded by the solution of the two single-objective optimisation problems. Again - and not surprisingly - both the gradient-based algorithm and the genetic algorithm outperform Sobol sampling. 

The widening of the gap between these methods relative to the case with three decision variables illustrates the curse of dimensionality in this optimisation problem. The design space grows exponentially with the number of decision variables. Both the sampling- and population-based methods must explore this space with finite number of evaluations. With 4096 Sobol samples in a six-dimensional space, the coverage is sparse and is equivalent to approximately four levels per variable if the sampling were uniform ($4^{6}=4096$). The gradient-based approach is less affected by this scaling as the gradient information available to the optimiser grows linearly with the number of decision variables, which provides directional information that guides the search. A further distinction between the three methods is in the termination criterion. The gradient-based optimisation provides a first-order optimality guarantee at convergence, terminating when the gradients of the objective and constraints satisfy the necessary conditions for a local optimum. The genetic algorithm and Sobol sampling, by contrast, are typically run with a predefined computational budget and offer no such guarantee. The limitation of the gradient-based approach is the potential for convergence to local optima, which can be mitigated through a multi-start strategy, though global optimality cannot be guaranteed.

\begin{figure}[h]
    \centering
    \includegraphics[width=1\linewidth]{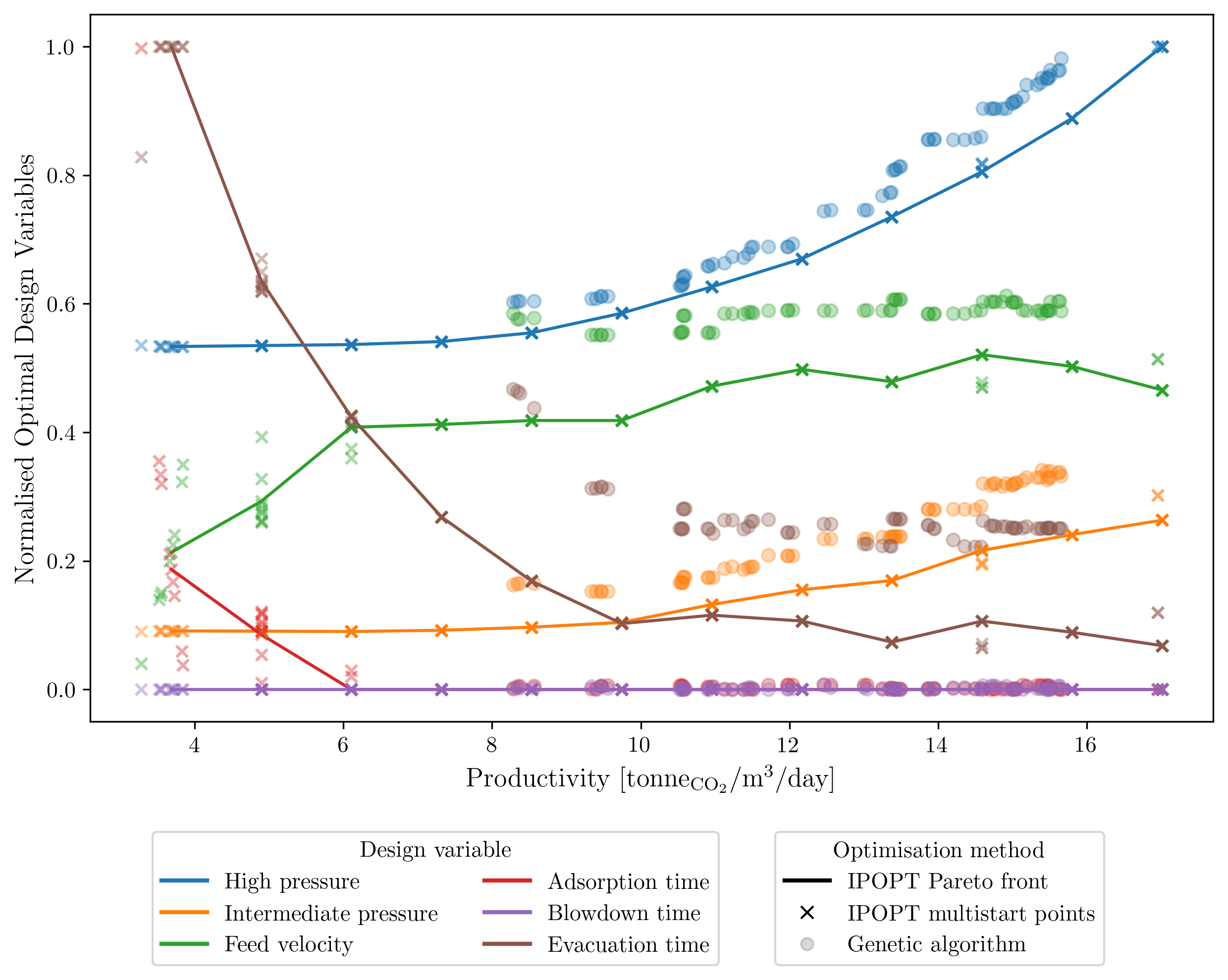}
    \caption{Decision variables of the optimal designs obtained by gradient-based optimisation (solids lines) and by the NSGA-II algorithm (symbols), i.e. the Pareto curves in Figure~\ref{constrained pareto front 6 variables}. The decision variables are normalised to their bounds and are plotted as a function of the productivity obtained for those points.}
    \label{dimensionless optimal variables for pareto}
\end{figure}

To further compare the solutions obtained with the gradient-based optimisation and the genetic algorithm, Figure \ref{dimensionless optimal variables for pareto} shows the values of the decision variables (normalised to their bounds) of the optimal designs as a function of the productivity (i.e. moving along the Pareto front). The solid lines refer to the solution of each epsilon-constrained sub-problem and the translucent crosses represent results from all 10 multi-start solutions for each sub-problem. The results from the NSGA-II algorithm are shown by the scattered translucent circles. Overall, there is general agreement between the trends outlined by the two optimisation algorithms. The shift of the results obtained with the genetic algorithm and their limited coverage of productivity values reflect the deficiencies observed in Figure~\ref{constrained pareto front 6 variables} when this algorithm is compared to the solution obtained by gradient-based optimisation. As discussed above, these differences reflect the lower precision of the population-based search within its computational budget.  

The results from the multi-start for the gradient-based optimisation provide further insight into the optimisation landscape. For the majority of the Pareto front, all 10 starts converge to the same solution in both the objective and decision variables space. At the low energy and low productivity end of the Pareto front, the optimisation landscape is very flat, as can be seen from the large range of productivities that yield a very similar energy consumption in Figure \ref{constrained pareto front 6 variables}. In this region, both the high pressure and the intermediate pressure converge to the same points. These variables contribute significantly to the energy consumption. The adsorption time and the feed velocity, however, freely trade off against each other, as their product determines the total feed volume per adsorption step. This observation suggests that the contribution of the pressure drop to the overall energy consumption is smaller than the optimality tolerance, and the optimiser does not resolve a unique combination of feed velocity and adsorption time. At the high productivity end of the Pareto front, local optima emerge. At the third highest productivity point on the front (14.6~$\mathrm{tonne}_{\mathrm{CO}_2}\,\mathrm{m}^{-3}\,\mathrm{day}^{-1}$), 7 of the 10 starts converge to the best solution, while 3 converge to a local optimum that differs by only 0.2~kWh/tonne in energy consumption. Yet, these solution differ significantly in their decision variables, suggesting that there are different arcs of optimal solutions within this region of the Pareto front. At the maximum productivity extreme, 2 of the 10 starts find a local optimum that is dominated by the other 8 solutions. In both scenarios where local optima are found, the impact on the objective function is small. 

\section{Discussion}\label{discussion}

\begin{figure}
    \centering
    \includegraphics[width=0.75\linewidth]{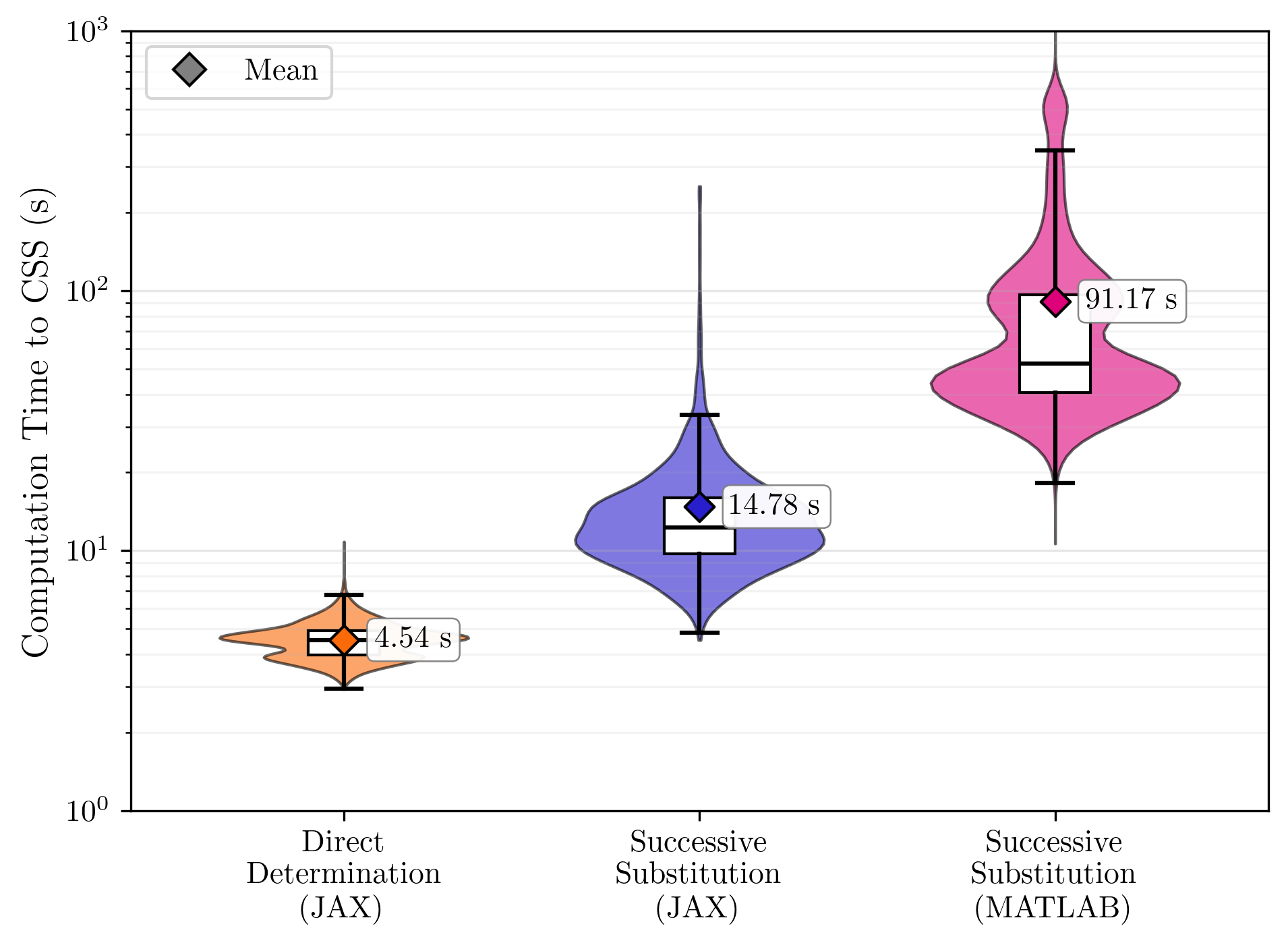}
    \caption{Violin plots of the computational time required to reach CSS for the direct determination and successive substitution methods in JAX as well as the successive substitution method in MATLAB across 4096 Sobol sampled designs. The white box illustrates the quartiles of the distribution and the whiskers correspond to 1.5 times the interquartile range. The mean of the distribution is indicated by the diamond symbol.}
    \label{violin plots computational time}
\end{figure}

The direct determination method requires significantly fewer iterations to attain CSS than successive substitution. However, each iteration of the direct determination method incurs additional computation for the evaluation of the cycle Jacobian (${\partial\textbf{x}_{k}}/{\partial\textbf{x}_{k}^{0}}$ in Equation \ref{error gradient}), such that a comparison of the total computational time is necessary. Furthermore, a comparison is made to the successive substitution method for a near-identical model implemented in MATLAB, which represents the standard practice for this model formulation. The two implementations differ in the solution method: the JAX formulation uses automatic differentiation for the Jacobian evaluation of the ODE right-hand side function within the implicit solver, rather than numerical approximation in the MATLAB version, and benefits from just-in-time compilation. As shown in Figure \ref{violin plots computational time} for 4096 Sobol sampled designs, the direct determination method in JAX requires an average wall-clock time of 4.54 seconds to reach CSS, compared to 14.78 seconds for successive substitution in JAX and approximately 20 times longer for the MATLAB implementation with successive substitution. These times do not include the compilation time for the JAX implementation, which is 26 seconds for the direct determination method. This is a one-time cost, as once compiled, all subsequent designs are evaluated without recompilation.

The gradient-based optimisation method outperforms both Sobol sampling and the NSGA-II algorithm in attaining the Pareto front for the multi-objective optimisation problems studied. Table~\ref{tab:optimisation_results} compares the number of model evaluations and the total computational time for each method applied to the productivity/energy usage design problem shown in Figures~\ref{3 variable optimisation constrained} and \ref{constrained pareto front 6 variables}. Again, we also compare in Table~\ref{tab:optimisation_results} solutions obtained using the differentiable model in JAX and the conventional model in MATLAB. The solution method in terms of the finite volume scheme, the ODE solver and the optimisation structure is the same for the model in JAX and the model in MATLAB, with the former providing computation of gradients to improve the computational speed and the Pareto front obtained. We acknowledge that the direct comparison of the computational time remain difficult as the methods differ both in their structure and in the quality of the solution. Yet, these values provide useful context for the practitioner. The Sobol sampling method evaluates 4096 points to attain both the maximal purity and recovery front and the productivity/energy Pareto front, unlike a targeted optimisation problem. The GA, with a population size of 72 and with 70 generations, evaluates the model 5040 times for a single optimisation problem. 

In the gradient-based optimisation, the problem is broken into single-objective sub-problems as well as having a multi-start to evaluate the non-convexity of the problem, such that the comparison is not like for like. For the purity/recovery optimisation problem, a total of 130 sub-problems are completed to obtain the pareto front with the multistart strategy. Each individual sub problem requires an average of 18 model evaluations and 188~s of wall-clock time to complete. The statistics for the productivity/energy consumption optimisation problems for both the three variable optimisation and the six variable optimisation are given in Table \ref{tab:optimisation_results}. For these problems, a total of 120 sub-problems are completed. An average of 13 model evaluations and a wall-clock time of 47 seconds are required to solve each sub-problem in the three dimensional optimisation, and 23 evaluations and 80 seconds for each sub-problem in the six variable optimisation. A multi-start strategy is not deemed essential for the productivity/energy consumption optimisation problem as the nonconvexity of the design space does not affect the obtained Pareto, thus it is not included in the statistics in Table \ref{tab:optimisation_results}. The statistics therefore just include 12 sub-problems for evaluating the Pareto front. As such, the gradient-based optimisation with the differentiable framework yields a speed-up of over 2 orders of magnitude compared to the state-of-the-art representative approach.

\begin{table}[h]
\centering
\begin{threeparttable}
\begin{tabular}{lrrrrr}
\toprule
& \multicolumn{4}{c}{\textbf{Differentiable Model (JAX)}} 
& \textbf{Conventional} \\
& \multicolumn{4}{c}{} 
& \textbf{Model (MATLAB)} \\
\cmidrule(lr){2-5}
& \textbf{IPOPT} & \textbf{IPOPT} & \textbf{NSGA-II} & \textbf{Sobol} & \textbf{NSGA-II} \\
& \textbf{(3 var.)} & \textbf{(6 var.)} & & \textbf{Sampling} & \\
\midrule
\textbf{Model evaluations}\tnote{a} & 167        & 315        & 5{,}040  & 4{,}096  & 5{,}040  \\[4pt]
\textbf{Wall-clock time}       & 9.5 min     & 16 min     & 7.9 hr   & 5.2 hr   & 129 hr   \\[4pt]
\textbf{Speedup}        &  & $\times$486 & $\times$16 &  & $\times$1 \\
\bottomrule
\end{tabular}
\begin{tablenotes}
    \footnotesize
    \item[a] IPOPT evaluation counts are for 12 Pareto front points without multistart
\end{tablenotes}
\end{threeparttable}
\caption{The performance of the differentiable model in JAX is evaluated across different  optimisation methods and against the conventional benchmark model.}
\label{tab:optimisation_results}
\end{table}

We note that parallelism offers a direct route to reducing the computational time of adsorption process optimisation, as many of the computational workflows discussed involve independent model evaluations that can be distributed across multiple processors. The extent to which each method benefits from parallelisation, however, differs. Sobol sampling is trivially parallel, as all 4096 designs are independent and can be evaluated simultaneously given sufficient computational resources. The NSGA-II genetic algorithm is parallel within each generation, where the population of candidate designs can be evaluated concurrently, though synchronisation is required between generations. Each individual optimisation run for the gradient-based optimisation is inherently sequential, however the parallelism potential comes with the potential to run each sub-problem in the multi-objective optimisation in parallel. Similarly, if a multi-start strategy is desired, these can be run in parallel. 

\section{Conclusion}\label{conclusion}

The design of cyclic adsorption processes presents a significant computational challenge. The requirement to simulate from startup to cyclic steady state for every candidate design, combined with the use of derivative-free optimisation, results in design campaigns requiring hundreds to thousands of CPU hours. This burden limits the ability to systematically explore the expanding design space of new adsorbent materials, cycle configurations, and operating conditions. In this work, we have demonstrated that differentiable programming addresses this challenge by transforming the entire process model into an end-to-end differentiable computational workflow, while retaining the same physical model and numerical discretisation used throughout the adsorption literature. A four-step pressure vacuum swing adsorption process for post-combustion carbon capture was implemented in JAX, with automatic differentiation providing exact gradients throughout the model without the manual derivation of sensitivity equations.
 
The framework delivers acceleration at three levels of the design workflow. At the level of the ODE solution, just-in-time compilation and automatic differentiation of the implicit solver reduce the simulation time to cyclic steady state from 91 seconds of wall-clock time in MATLAB to 15 seconds in JAX for the successive substitution method. At the level of cyclic steady state convergence, the differentiation of a single process cycle provides the Jacobian required for a Newton-based direct determination method. This exploits the near-quadratic convergence of the Newton iteration, which requires an average of 5.1 iterations to reach cyclic steady state compared to 112 for successive substitution, further reducing the simulation time to 4.5 seconds --- a reduction of approximately 20-fold relative to the conventional MATLAB implementation. At the level of the optimisation, the availability of exact gradients of the key performance indicators with respect to the design variables enables the use of the IPOPT algorithm for multi-objective optimisation via the epsilon-constraint method. For the six-variable problem maximising productivity and minimising energy consumption subject to purity and recovery constraints, the gradient-based approach obtains the Pareto front in approximately 16 minutes --- over two orders of magnitude faster than the conventional genetic algorithm approach. The resulting front is also of superior quality, capturing the extremes of the trade-off curve that the genetic algorithm does not reach.
 
The central advantage of the approach lies in its generality. Because automatic differentiation operates throughout the computational workflow irrespective of the specific model formulation, exact gradients are made available wherever they are useful, without the manual derivation of sensitivity equations that has historically limited the adoption of gradient-based methods in the adsorption community. The framework is therefore readily applicable to alternative adsorbents, cycle configurations, and column designs. Future work will extend the approach to more complex multi-column cycles, integrate the differentiable model with GPU-accelerated solvers for the massively parallel evaluation of designs, and exploit the available gradients for uncertainty quantification and adsorbent screening. More broadly, this work illustrates how differentiable programming can shift cyclic process design from slow black-box simulation with derivative-free optimisation towards efficient, gradient-enhanced modelling and optimisation, with relevance to the wider class of periodic processes encountered in chemical engineering.

\newpage
\section{Acknowledgements}

A.G. gratefully acknowledges funding from the EPSRC DTP scholarship granted to the Department of Chemical Engineering at Imperial College London and by the Bansal Bursary. 
The authors also thank Ben Moseley and Ashwin Kumar Rajagopalan, along with members of their respective research groups, for valuable discussions.

\bibliography{references}

\end{document}